\documentclass[twocolumn]{article}

\usepackage{abstract}
\usepackage{adjustbox}
\usepackage{algorithm}
\usepackage{amsmath}
\usepackage{amsfonts} 
\usepackage{color}
\usepackage[margin=0.6in]{geometry}
\usepackage{graphicx}
\usepackage{hyperref}
\usepackage[T1]{fontenc}
\usepackage{textcomp}
\usepackage{listings}
\usepackage{mathtools} 
\usepackage{multirow}
\usepackage{siunitx}  

\usepackage{subfig}
\usepackage{xspace}

\let\qty\SI
\let\qtylist\SIlist

\definecolor{textblue}{rgb}{.2,.2,.7}
\definecolor{textred}{rgb}{0.54,0,0}
\definecolor{textgreen}{rgb}{0,0.43,0}
\definecolor{textpurple}{rgb}{0.59,0,1}
\definecolor{textcomment}{rgb}{0.2,0.43,0.43}

\newcommand*\samethanks[1][\value{footnote}]{\footnotemark[#1]}
\newcommand{\pygbe}{\texttt{PyGBe}\xspace}
\newcommand{\gmres}{\textsc{gmres}\xspace}
\newcommand{\bem}{\textsc{bem}\xspace}
\newcommand{\fmm}{\textsc{fmm}\xspace}
\newcommand{\kifmm}{\textsc{kifmm}\xspace}
\newcommand{\ncrit}{n_{\mathrm{crit}}}  

\newcommand{\ie}{\textit{i}.\textit{e}., }
\DeclareSIUnit\angstrom{\text{Å}}       

\graphicspath{{figs/}}

\title{High-productivity, high-performance workflow for virus-scale electrostatic simulations with Bempp-Exafmm}

\author{%
    Tingyu Wang\thanks{Department of Mechanical and Aerospace Engineering, The George Washington University, Washington, DC, USA}%
    \and Christopher D. Cooper\thanks{Department of Mechanical Engineering and Centro Cient\'ifico Tecnol\'ogico de Valpara\'iso, Universidad T\'ecnica Federico Santa Mar\'ia, Valpara\'iso, Chile}%
    \and Timo Betcke\thanks{Department of Mathematics, University College London, UK}%
    \and Lorena A. Barba\samethanks[1]%
}

\date{}

\begin{document}

\twocolumn[
\maketitle

\begin{onecolabstract}
Biomolecular electrostatics is key in protein function and the chemical processes affecting it.
Implicit-solvent models via the Poisson-Boltzmann (PB) equation provide insights with less computational cost than atomistic models, making large-system studies---at the scale of viruses---accessible to more researchers.
    Here we present a high-productivity and high-performance linear PB solver based on Exafmm, a fast multipole method library, and Bempp, a Galerkin boundary element method package.
    The workflow integrates an easy-to-use Python interface with optimized computational kernels, and
    can be run interactively via Jupyter notebooks, for faster prototyping.
    Our results show the capability of the software, confirm code correctness, and assess performance with between 8,000 and 2 million elements.
    Showcasing the power of this interactive computing platform, we study the conditioning of two variants of the boundary integral formulation with just a few lines of code.
    Mesh-refinement studies confirm convergence as $1/N$, for $N$ boundary elements, and
    a comparison with results from the trusted APBS code using various proteins shows agreement.
    Our binding energy calculations using 9 various complexes align with the results from using five other grid-based PB solvers.
    Performance results include timings, breakdowns, and computational complexity.
    Exafmm offers evaluation speeds of just a few seconds for tens of millions of points, and $\mathcal{O}(N)$ scaling.
    The trend observed in our performance comparison with APBS demonstrates the advantage of Bempp-Exafmm in applications involving larger structures or requiring higher accuracy.
    Computing the solvation free energy of a Zika virus, represented by 1.6 million atoms and 10 million boundary elements, took 80-min runtime on a single compute node (dual 20-core).

\end{onecolabstract}
]

\section{Introduction}\label{sec:intro}
Electrostatics plays a key role in the structure and function of biological molecules.
Long-range electrostatic effects intervene in various essential processes, such as protein binding, with biomolecules always present in a solution of water with ions.
Computer simulations to study electrostatic interactions in biomolecular systems divide into those that represent the solvent explicitly---in full atomic detail---or implicitly.
In so-called implicit-solvent models~\cite{RouxSimonson1999,DecherchiETal2015}, the solvent degrees of freedom are averaged out in a continuum description.
Starting from electrostatic theory, this leads to a mathematical model based on the linearized Poisson-Boltzmann equation, and widely used to compute mean-field electrostatic potentials and solvation free energies.
Poisson-Boltzmann solvers have been numerically implemented using finite difference \cite{RocchiaAlexovHonig2001, BakerETal2001, chen2011mibpb}, finite element \cite{BakerETal2001,BondETal2010,HolstETal2012}, boundary element \cite{AltmanBardhanWhiteTidor2009, GengKrasny2013, ZhangPengHuangPitsianisSunLu2015, CooperBardhanBarba2014}, and (semi) analytical \cite{LotanHead-Gordon2006,FelbergETal2017} methods, scaling up to problems as large as virus capsids \cite{ZhangETal2019,MartinezETal2019}.

Virus-scale simulations are at the limit of what can be accomplished in computational biophysics, using leadership computing facilities.
These computational studies provide fundamental insight to understand the physical underpinnings of virus particles \cite{HaddenPerilla2018}, such as their mechanical properties \cite{ArkhipovETal2009}, binding mechanisms \cite{DurrantETal2020}, assembly \cite{DickETal2018}, and structure \cite{TurovnovaETal2020}, among others. 
The first explicit-solvent atomic simulation of a virus using molecular dynamics was published just 15 years ago, modeling a plant virus (satellite tobacco mosaic virus) of 1.7 nm in diameter \cite{FreddolinoETal2006}.
The full model included 1 million atoms, and the computations ran for many days on the world-class facilities at the National Center for Supercomputing Application (NCSA), University of Illinois.
Using largely the same methods, researchers just last year could model the full viral envelope of a 2009 pandemic influenza A H1N1 virus, with a diameter of about 115 nm \cite{DurrantETal2020}.
In this case, the full system consisted of 160 million atoms, and the computations ran on the Blue Waters supercomputer at NCSA using 115k processor cores (4,096 physical nodes).
This is among the largest biomolecular systems ever simulated using all-atom molecular dynamics.

Only a few elite researchers can access these leadership computing facilities, however, and if molecular science of viruses is to progress, computational tools that are more widely accessible are needed.
It is in this context where approximate coarse grained \cite{ReddySansom2016,MachadoETal2017,HuberETal2021} and continuum models \cite{MartinezETal2019} play a key role.
The vision behind this paper is to build an electrostatic simulation platform for biomolecular applications that allows researchers to access it via the Python/Jupyter ecosystem. This provides a high degree of flexibility in the underlying formulations, rapid prototyping of novel models, ease of deployment and integration into existing simulation workflows.

To achieve this vision, we are coupling two libraries, the high-level Galerkin boundary element library Bempp, which is fully developed in Python, and the very fast low-level high-performance fast multipole method (\fmm) library Exafmm. 
Boundary integral problems are described in Bempp using a high-level approach that enables building even complex block-operator systems in just a few lines of Python code. Bempp then executes the discretization, depending on the chosen parameters and machine environment. 
Exafmm is called as a matrix-vector black-box below the user level, hiding all technicalities associated with the discretization.

This approach has the following advantages as compared to an integrated PB solver implemented in, for example, C++:
\begin{itemize}
	\item \textit{Strict separation of concerns}. The user-level description of the electrostatic problem is completely separated from the underlying discretization routines and the \fmm coupling. One can easily move between different types of implementations (e.g., dense discretization, \fmm) editing a single parameter, change input file handling or postprocessing.
	\item \textit{Fast prototyping of different formulations}. We present in this paper results produced with a direct formulation and derivative or Juffer-type formulations. Applying these different formulations requires editing just a few lines of high-level code. The user can easily experiment with other models, such as piecewise solvation models with different solvation parameters in each layer.
	\item \textit{Portability}. Bempp and Exafmm can easily be installed as a joint Docker image that is automatically tracking the current development of these libraries. The whole solution workflow can be implemented in a brief Jupyter notebook.
\end{itemize}
A high-level productive approach does come with some costs. A dedicated highly specialized C++ code that integrates all steps might be faster than our solution. Nevertheless, in this paper we demonstrate that our software platform is highly competitive for real-world solvation energy computations (and many other electrostatic computations), while preserving full flexibility through the use of a high-productivity Python environment.

We present results that show the power of interactive computing to study modeling variations, results to confirm code correctness and describe performance, and a final showcase that computes solvation free energy for a medium-sized virus particle.
Our first result explains the behavior of two solution methods that vary in whether they solve for the potential internal or external to the molecular interface, from the conditioning point of view.
Solution verification via grid-convergence studies with two problem set-ups and the comparison with trusted community software give us confidence in the software implementation.
Performance-wise, we show results with problem sizes up to 2 million boundary elements, we show computational complexity of the \fmm evaluations, and timing breakdowns of the solver.
We also compare our performance with APBS using a moderate-sized protein under various levels of accuracy.
The final result uses the enveloped Zika virus, computing the surface potential and solvation free energy with about 10 million boundary elements.
All results are reproducible and we share scripts, data, configuration files, and Jupyter notebooks in the manuscript repository, found at \url{https://github.com/barbagroup/bempp_exafmm_paper/}, in addition to permanent archives in Zenodo.
Permanent identifiers are provided at the end of the Results section.

\section{Results}\label{sec:results}
We demonstrate the performance and capability of Bempp-Exafmm via electrostatic simulations, including computing the solvation energy of a Zika virus.
Bempp solves the boundary integral formulation of the linearized PB equation.
This section presents six types of results.
The first result explains the behavior of two variants of the mathematical formulation, from the conditioning point of view. 
Second, we show solution verification through two grid-convergence studies: with a spherical molecule (having an analytical solution), and with a real biomolecule (using Richardson extrapolation).
Third, we provide more evidence of our software correctness via binding energy computations.
Fourth, we demonstrate the performance with problem sizes between 8,000 and 2 million elements, including timings, breakdowns, and computational complexity.
The fifth type of result compares the performance of Bempp and APBS across different accuracy using a medium-sized protein.
Our final result is a demonstration using a structure with about 1.6 million atoms, the Zika virus, discretized with about 10 million boundary elements.
In the supplementary material, we show agreement with APBS by computing the solvation energy of 9 different molecules of varying sizes.

We ran all experiments on a single CPU node of \textit{Pegasus}, a Linux cluster at the George Washington University.
Each node is equipped with two 20-core Intel Xeon Gold 6148 CPUs (base frequency at 2.4 GHz, max turbo frequency at 3.7 GHz) and 192GB RAM.
All runs are based on Bempp-cl version 0.2.2 and Exafmm-t version 0.1.0.
We compiled Exafmm with Intel compiler (version 19.0.5.281) and enabled \texttt{-xHost} option for vectorization.
We used the full GMRES from the SciPy library as our linear solver.

In all of the following test cases, the relative permittivity is $\epsilon_1 = 4$ in the solute region and $\epsilon_2 = 80$ in the solvent region, and the inverse of Debye length in the solvent region is $\kappa = \qty{0.125}{\angstrom}^{-1}$, corresponding to a salt concentration of 150 mM.
We downloaded the molecule structures from the Protein Data Bank (PDB), parameterized them with \texttt{pdb2pqr}~\cite{DolinskyETal2004}, and generated meshes on the solvent-excluded surface (SES) using \texttt{Nanoshaper} with a probe radius of $\qty{1.4}{\angstrom}$.

\paragraph{Matrix conditioning of two derivative formulations} \label{result_conditioning}
Section \ref{s:formulation} presents the two formulations to solve the integral equations  \eqref{eq:volume_potential} derived from the Poisson-Boltzmann model of biomolecular electrostatics: 
the \emph{direct}~\cite{YoonLenhoff1990}  and \emph{derivative}~\cite{JufferETal1991} formulations.
The latter is well-known to lead to a better-conditioned matrix.
Its most common solution method finds the potential and its derivative in the interior of the boundary via equations \eqref{eq:juffer}, but an alternative is to solve for the exterior fields via \eqref{eq:lu}.
Bempp unfetters the user to experiment with these variants of the boundary element solution method, editing just a few lines of Python.

Bempp allows users to define function spaces and integral operators effortlessly via high-level interfaces.
For example, we can define a space of continuous piecewise linear functions on the mesh and then create a Laplace single-layer potential boundary operator using the code shown in Listing \ref{lst:space_and_operator}.
The variable \texttt{grid} represents the surface mesh.
\begin{adjustbox}{max width=\linewidth}
    \begin{lstlisting}[caption=Define function spaces and boundary operators in Bempp.,
        label={lst:space_and_operator}]
from bempp.api.operators.boundary import laplace
space = bempp.api.function_space(grid, "P", 1)
slp_l = laplace.single_layer(space, space,
    space, assembler="fmm")
    \end{lstlisting}
\end{adjustbox} \\

\noindent
Other boundary operators can be defined in a similar fashion.
Setting up the system matrix of the two derivative formulations, corresponding to Equation \ref{eq:juffer} and \ref{eq:lu}, is straightforward, as shown in the code snippets below.

\begin{adjustbox}{max width=\linewidth}
    \begin{lstlisting}[caption=Interior formulation.,
        label={lst:interior_formulation},
        emph={id, slp_l, slp_y, dlp_l, dlp_y,
        hyp_l, hyp_y, adj_l, adj_y},
        emphstyle=\color{textblue}
        ]
ep = ep_ex/ep_in   # ratio of permittivity
A = bempp.api.BlockedOperator(2,2)
A[0,0] = 0.5*(1+ep)*id - (ep*dlp_y - dlp_l)
A[0,1] = slp_y - slp_l
A[1,0] = hyp_y - hyp_l
A[1,1] = 0.5*(1+1/ep)*id + (1/ep)*adj_y - adj_l
    \end{lstlisting}
\end{adjustbox}

\begin{adjustbox}{max width=\linewidth}
    \begin{lstlisting}[caption=Exterior formulation.,
        label={lst:exterior_formulation},
        emph={id, slp_l, slp_y, dlp_l, dlp_y,
        hyp_l, hyp_y, adj_l, adj_y},
        emphstyle=\color{textblue}
        ]
ep = ep_ex/ep_in
A = bempp.api.BlockedOperator(2,2)
A[0,0] = 0.5*(1+1/ep)*id - (dlp_y - 1/ep*dlp_l)
A[0,1] = slp_y - slp_l
A[1,0] = 1/ep*(hyp_y - hyp_l)
A[1,1] = 0.5*(1+1/ep)*id + (1/ep)*adj_y - adj_l
    \end{lstlisting}
\end{adjustbox} \\

\noindent
\texttt{id} is the identity operator, and other variables in blue are boundary operators.
The prefix denotes the type of operators (single-layer, double-layer, hypersingular or adjoint double-layer), while the suffix suggests whether the operator has a Laplace kernel (in the Laplace equation) or Yukawa kernel (in the linearized PB equation).
For example, the variable \texttt{hyp\_y} is a Yukawa hypersingular boundary operator, corresponding to the term $W_{Y}^{\Gamma}$ in Equation \ref{eq:juffer} and \ref{eq:lu}.

The typical workflow for solving PB equations using BEM software involves the following steps: obtain structures from PDB, parameterize with \texttt{pdb2pqr}, generate the surface mesh, run linear solvers and compute energies.
What differentiates Bempp-Exafmm from other ``black-box'' PB solvers is the ability to customize each step---such as the formulation and preconditioner---in the computation, making it a great testbed for exploratory numerical analysis.
We could easily try both the interior and exterior versions of the derivative formulation, whereas previous publications opted for one method and used it throughout.
The programming effort required to implement a second formulation would have been a good reason.
In our experiments, the exterior version used by Lu and coworkers~\cite{LuETal2006,LuETal2009,ZhangETal2019} took about half as many iterations to converge than the interior version---a sizable advantage.
This led us to study the properties of the two variants of the derivative formulation in more detail.
The results in this section aim to give a simple explanation for the different numerical behavior of the two methods.
Our ability to explore and explain this issue showcases the power of interactive computing with a high-productivity software platform, like that provided by Bempp-Exafmm.

\begin{figure*}
    \begin{center}
        \subfloat[][]{\includegraphics[width=0.4 \textwidth]{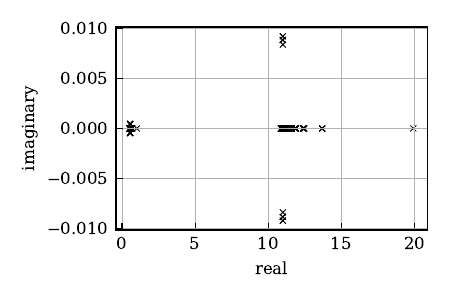}
        \label{fig:derivative_interior_eig}}\qquad
        \subfloat{\includegraphics[width=0.4 \textwidth]{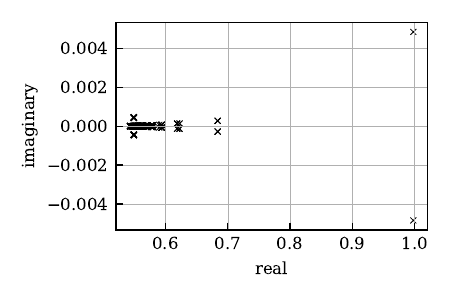}
        \label{fig:derivative_exterior_eig}}
    \end{center}
    \caption{Eigenvalues of the system matrix of the derivative formulation for interior field (\textbf{a}) and for exterior field (\textbf{b}).
    }
\end{figure*}

GMRES methods have an intricate convergence behavior \cite{mark1999a}.
Heuristically, if the eigenvalues are clustered with the cluster being sufficiently far away from the origin, we expect fast convergence of GMRES to the desired solution.
Figures \ref{fig:derivative_interior_eig} and \ref{fig:derivative_exterior_eig} show the eigenvalues of the interior and exterior derivative formulations, respectively, on the complex plane.
With the interior formulation, eigenvalues cluster around two points, while eigenvalues cluster around only one point with the exterior formulation.

The difference is due to the diagonal of the corresponding system of integral equations.
In the case of the interior formulation, the associated left-hand side operator takes the form
$$
\begin{bmatrix}\frac{1}{2}(1 + \frac{\epsilon_2}{\epsilon_1})I & 0 \\ 0 & \frac{1}{2}(1 + \frac{\epsilon_1}{\epsilon_2})I
\end{bmatrix} + \mathcal{C}_{int},
$$
where $\mathcal{C}_{int}$ is a compact operator on sufficiently smooth domains. (On smooth domains the single-layer, double-layer and adjoint double-layer operators are compact operators.
Furthermore, the difference of the hypersingular operators is compact \cite{Hiptmair2006-om}.)
The eigenvalues of the interior derivative operator hence accumulate at the points $\frac{1}{2}(1 + \frac{\epsilon_2}{\epsilon_1})$ and $\frac{1}{2}(1 + \frac{\epsilon_1}{\epsilon_2})$.
In contrast, the exterior derivative operator has the form
$$
\begin{bmatrix}\frac{1}{2}(1 + \frac{\epsilon_1}{\epsilon_2})I & 0 \\ 0 & \frac{1}{2}(1 + \frac{\epsilon_1}{\epsilon_2})I
\end{bmatrix} + \mathcal{C}_{ext},
$$
where $\mathcal{C}_{ext}$ is again a compact operator.
We now only have one accumulation point, namely $\frac{1}{2}(1 + \frac{\epsilon_1}{\epsilon_2})$.
Unless $\epsilon_1\approx \epsilon_2$ we therefore expect the eigenvalues to be much closer together than in the interior derivative case and therefore the GMRES convergence to be faster for the exterior derivative formulation.

We want to emphasize that the above argument is valid for the continuous operators.
Under discretization, the resulting eigenvalue problem is of the form $A\mathbf{x}=\lambda M\mathbf{x}$ (or equivalently $M^{-1}A\mathbf{x}=\lambda \mathbf{x}$, where $A$ is the $2\times 2$ block operator associated with the Galerkin discretization of the integral operator system and $M = \text{diag}(\hat{M}, \hat{M})$ is a block diagonal mass matrix, where the matrix $\hat{M}$ is the matrix containing the surface inner products $\int_{\Gamma}\psi_i(\mathbf{r})\phi_j(\mathbf{r})ds(\mathbf{r})$ for test functions $\psi_j$ and trial functions $\phi_i$ (which are both chosen as continuous, piecewise linear basis functions for the derivative formulation in this paper).
The mass-matrix preconditioned linear system of equations to solve has the form $M^{-1}A\mathbf{x} = M^{-1}\mathbf{b}$ for vector of unknowns $\mathbf{x}$ and right-hand side $\mathbf{b}$ and Figures \ref{fig:derivative_interior_eig} and \ref{fig:derivative_exterior_eig} show the eigenvalues of $M^{-1}A$ for the interior and exterior derivative formulation.
In practice, the action of $M^{-1}$ can be computed through a sparse LU decomposition of $M$.
However, for problems with millions of unknowns this is becoming expensive.
For practical electrostatic computations we have therefore chosen a simple mass lumping approach, in which we substitute $M$ by a diagonal matrix, where each diagonal entry is the sum of the corresponding row values of $M$.
This diagonal matrix can then be trivially inverted.
In our experiments the mass lumping only led to a modest increase in the number of iterations compared to using the LU decomposition of the mass matrix $M$.
We stress that mass matrix preconditioning is necessary to solve the interior and exterior formulations in a reasonable number of iterations.
The approximation of the inverse of the mass matrix $M$ through mass lumping only affects the preconditioner and not the solution of the underlying linear system of equations itself.

In each of the following studies, we present two sets of results: one from using the exterior derivative formulation with piecewise linear elements, preconditioned by a mass lumping matrix, and the other from using the direct formulation with piecewise constant elements, preconditioned by a block-diagonal matrix presented by Altman and co-workers \cite{AltmanBardhanWhiteTidor2009}.

\paragraph{Mesh refinement study using an analytical solution} \label{result_convergence_sphere}

\begin{figure*}
        \centering
     \subfloat[][Sphere with an off-center unit charge at $(1,1,1)$.]{\includegraphics[width=0.4 \textwidth]{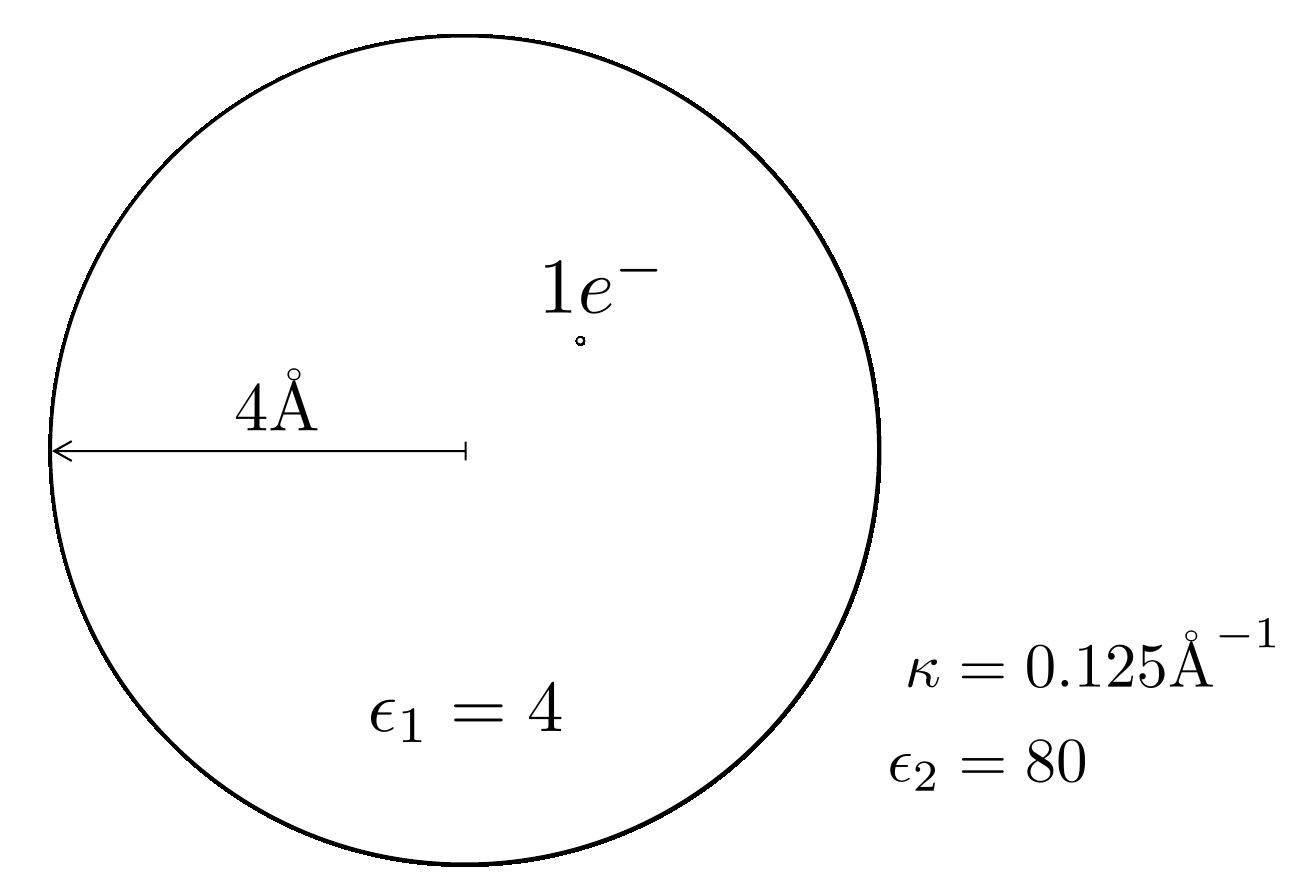}
        \label{fig:sketch_sphere_convergence}}\quad
     \subfloat[][Structure of 5PTI.]{\includegraphics[width=0.35 \textwidth]{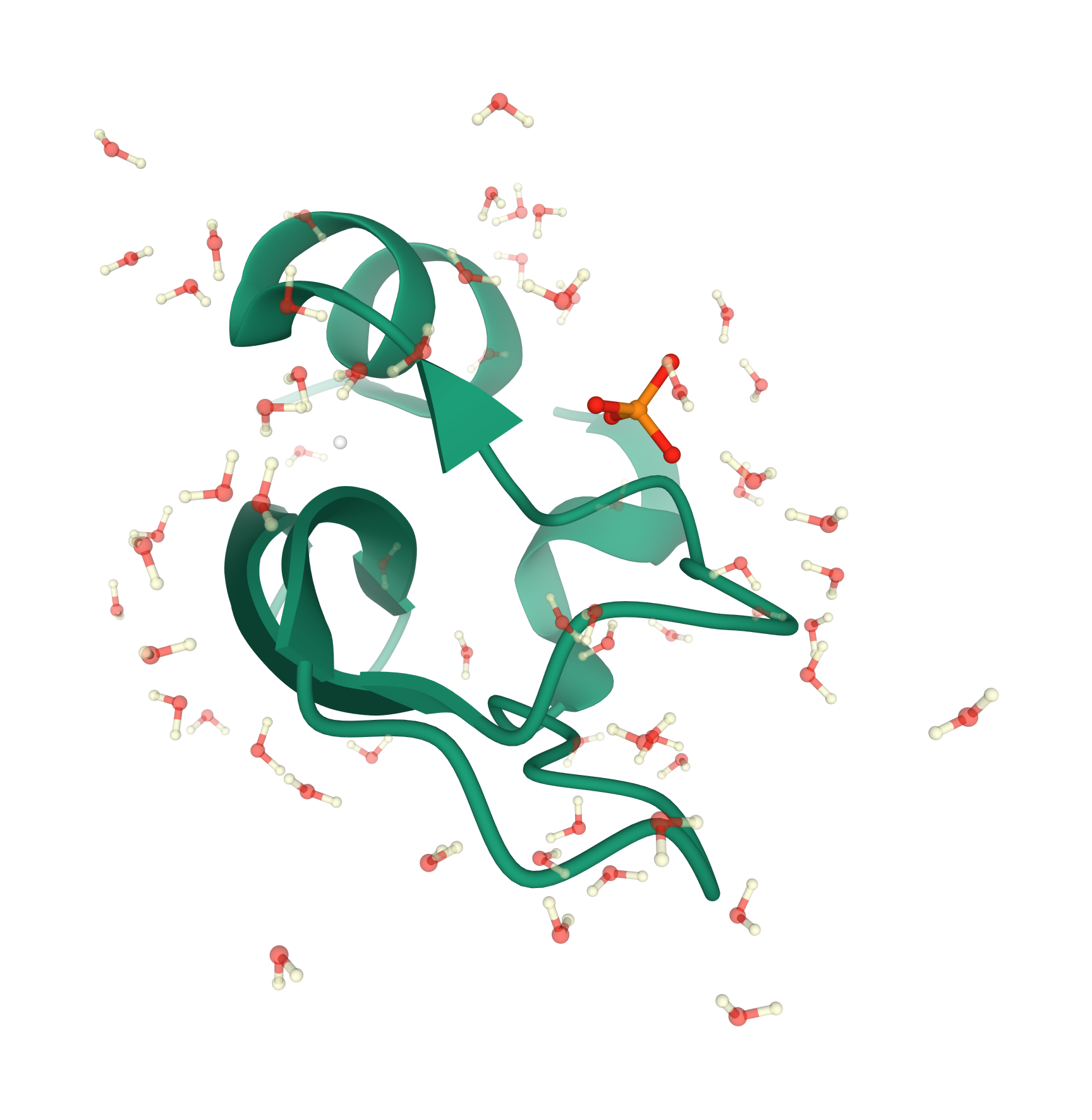}
        \label{fig:5PTI_structure}}\\
     \subfloat[][]{\includegraphics[width=0.4 \textwidth]{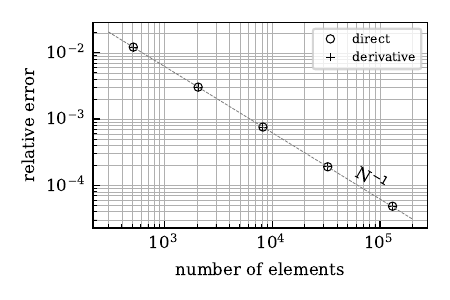}
        \label{fig:sphere_convergence}}
     \subfloat[][]{\includegraphics[width=0.4 \textwidth]{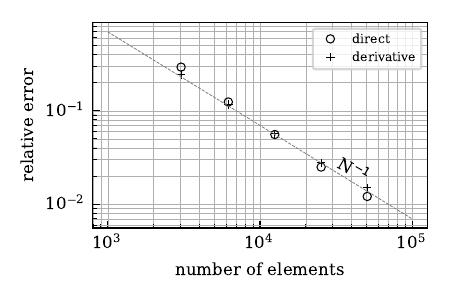}
        \label{fig:5PTI_convergence}}
    \caption{Mesh refinement studies using a spherical molecule and a real biomolecule: bovine pancreatic trypsin inhibitor (PDB code 5PTI).
    \textbf{c}, Mesh convergence study on a spherical molecule with an off-center charge, using both direct formulation and derivative formulation. The error on the solvation energy with respect to the analytical solution is plotted for five meshes:
    the sphere discretized with 512, 2048, 8192, 32768 and 131072 boundary elements.
    \textbf{d}, Mesh convergence study of the solvation energy of bovine pancreatic trypsin inhibitor (PDB code 5PTI), using both direct formulation and derivative formulation.
    The error is with respect to the extrapolated solution using Richardson extrapolation.
    }
\end{figure*}

As a form of solution verification with the Bempp-Exafmm software, we completed two mesh-refinement studies.
The first used a spherical molecule with an off-center charge, for which we have an analytical solution.
In the next sub-section, we present a mesh-refinement study with a real molecule of biological relevance.
Figure \ref{fig:sketch_sphere_convergence} depicts the problem setup for the current case:
a spherical molecule of radius \qty{4}{\angstrom} and relative permittivity $\epsilon_1 = 4$, with a unit charge located at $(1,1,1)$.
The solvent region has the relative permittivity of water ($\epsilon_2 = 80$), and a salt concentration of $150$mM $(\kappa = \qty{0.125}{\angstrom}^{-1})$.
Other simulation parameters are listed in Table \ref{tab:convergence}.
With an expansion order of 10, our \fmm achieved 9 digits of accuracy.
We computed the solvation energy of this molecule using $5$ different meshes, obtained using a constant refinement factor of 4.

\begin{table}[]
    \centering
    \begin{tabular}{lc}
    \hline
    \gmres tolerance          & $10^{-7}$ \\
    \# regular quadrature points  & 6    \\
    \fmm expansion order      & 10   \\
    \fmm $\ncrit$             & 500  \\
    \hline  \vspace{0.3 cm}
    \end{tabular}

    \begin{tabular}{cc}
    number of elements & mesh density ($\#/{\si{\angstrom}}^2$) \\
    \hline
    3032               & 1                                       \\
    6196               & 2                                       \\
    12512              & 4                                       \\
    25204              & 8                                       \\
    50596              & 16                                     
    \end{tabular}
    \caption{Simulation parameters for the grid-refinement studies (top); mesh sizes/densities used in the grid refinement study on 5PTI. Mesh densities measured as number of elements per square Angstrom.}
    \label{tab:convergence}
\end{table}

Kirkwood's derivation \cite{kirkwood1934theory} allows us to compute the analytical solution for the solvation energy in this case, to compare with the numerical result: $-12.258363$ [kcal/mol].
Figure \ref{fig:sphere_convergence} shows the error in the solvation energy, converging at the expected rate of $1/N$ for both formulations.
The observed order of convergence is 1.001 for the direct formulation and 0.999 for the derivative formulation, using the middle three values.

\paragraph{Mesh refinement study using 5PTI} \label{result_convergence_5PTI}

Next, we tested our software using a real biomolecule: bovine pancreatic trypsin inhibitor (PDB code 5PTI), whose structure \cite{wlodawer1984structure} is shown in Figure \ref{fig:5PTI_structure}.
We parameterized the molecule with \texttt{pdb2pqr} and the \texttt{charmm} \cite{brooksCHARMMProgramMacromolecular1983} force field, and then computed the solvation energy using 5 meshes with the element density ranging from 1 to 16 (Table \ref{tab:convergence}).
This test used the same parameters listed in Table \ref{tab:convergence}, which are fine enough to reveal the discretization error.
Since an analytical solution is not available for this geometry, we obtained the reference values for error estimation via Richardson extrapolation.
The estimated relative error with the finest mesh is 1.2\% with the direct formulation, and 1.5\% with the derivative formulation.

Figure \ref{fig:5PTI_convergence} shows that the error of the computed solvation energy for 5PTI converges linearly with respect to $N$.
The observed order of convergence is 1.156 for the direct formulation and 1.038 for the derivative formulation, using the middle three values.
Both convergence results provide solution verification, and are evidence that our software solves the mathematical model correctly.

\paragraph{Binding energy calculations}

Computing the binding free energy between two molecules is a more challenging test for code verification. 
This quantity is computed as the energetic difference between the bound complex and the two isolated compounds:
\begin{equation} \label{eqn:bind}
\Delta G^{polar}_{bind} = \Delta G^{polar}_{solv,complex} - \Delta G^{polar}_{solv,1} - \Delta G^{polar}_{solv,2} - \Delta G_{coul} 
\end{equation}
where $\Delta G_{coul}$ is the difference in Coulombic energy between the bound and unbound states. 
As $\Delta G^{polar}_{bind}$ is a small difference between large solvation energies, accurate calculations are key for reliable results. 
Binding energy calculations with finite-difference codes were thoroughly studied by other researchers for a large test set (51 compounds) with several solvers, in particular, adaptive cartesian grid-based PB solver (CPB) \cite{HarrisBoschitcshFenley2013}, DELPHI, MIBPB, PBSA, and APBS \cite{nguyenAccurateRobustReliable2017}.
In these tests, the authors were specially careful regarding mesh refinement and accuracy.

\begin{figure*}
    \centering
    \includegraphics[width=0.8\textwidth]{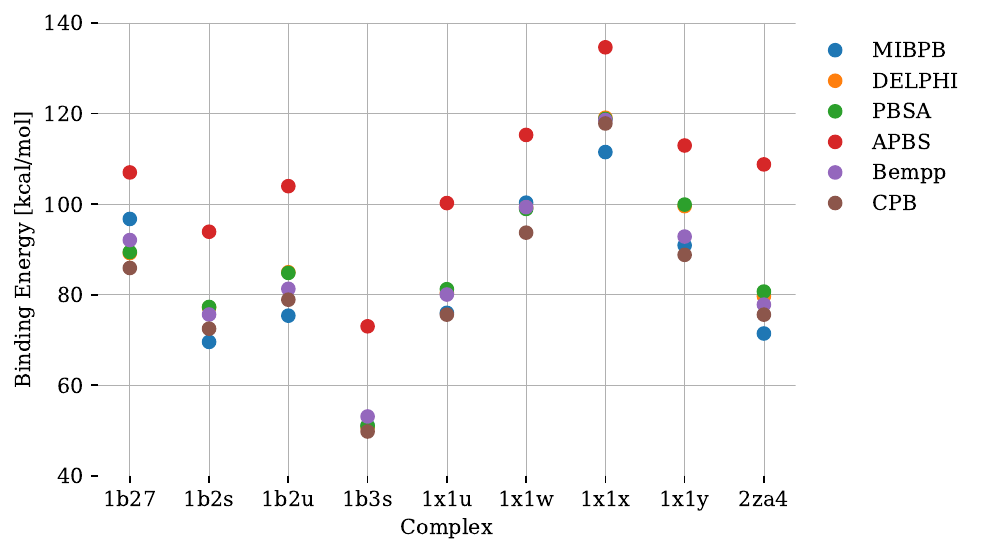}
    \caption{Binding energies of 9 Barnase-Barstar complexes computed using various PB solvers.
    }
    \label{fig:bind}
\end{figure*}

We performed the same test with Bempp to compute $\Delta G^{polar}_{bind}$ for a subset of the 51 compounds.
The pqr files were downloaded from \url{http://www.sb.fsu.edu/~mfenley/convergence/downloads/convergence_pqr_sets.tar.gz}.
The dielectric constants are 1 and 80 in the solute and solvent region, respectively, and $\kappa$ is set to 0.10265 (the ionic strength is 100 mM).
Table \ref{tab:bind} shows the binding energies of 9 Barnase-Barstar complexes computed with Bempp using the finest meshes (with a mesh density of 48 elements per $\si{\angstrom}^2$), compared with the finest mesh results reported by Harris \emph{et al.} \cite{HarrisBoschitcshFenley2013} and Nguyen \emph{et al.} \cite{nguyenAccurateRobustReliable2017}.
Figure \ref{fig:bind} demonstrates that our results are within the variability between other codes, indicating the correctness of our implementation.
The values of each component in Equation \ref{eqn:bind} are available in the supplementary material.

\begin{table*}[]
    \centering
    \begin{tabular}{c|c|c|c|c|c|c}
         & MIBPB & DELPHI & PBSA  & APBS  & Bempp & CPB \\ \hline
    1b27 & 96.7  & 89.2   & 89.4  & 107.0 & 92.1  & 85.9   \\
    1b2s & 69.6  & 77.2   & 77.3  & 93.9  & 75.6  & 72.5   \\
    1b2u & 75.4  & 85.0   & 84.8  & 104.0 & 81.3  & 78.9   \\
    1b3s & 51.1  & 50.7   & 51.1  & 73.0  & 53.1  & 49.8   \\
    1x1u & 76.0  & 80.4   & 81.2  & 100.2 & 80.0  & 75.6   \\
    1x1w & 100.3 & 99.3   & 98.9  & 115.3 & 99.3  & 93.7   \\
    1x1x & 111.5 & 119.1  & 118.8 & 134.6 & 118.6 & 117.8  \\
    1x1y & 90.9  & 99.5   & 99.9  & 112.9 & 92.8  & 88.8   \\
    2za4 & 71.4  & 79.7   & 80.7  & 108.8 & 77.8  & 75.6  
    \end{tabular}
    \caption{Binding free energies $\Delta G^{polar}_{bind}$ of the Barnase-Barstar complexes using different PB solvers.
    Energies are in units of kcal/mol. The first 4 columns were taken from the work by Nguyen \emph{et al.} \cite{nguyenAccurateRobustReliable2017} and the last column from Harris \emph{et al.} \cite{HarrisBoschitcshFenley2013} }
    \label{tab:bind}
\end{table*}

\paragraph{Performance study with direct and derivative formulations} \label{result_performance}

In this sub-section, we investigate the computational performance of Bempp-Exafmm using 1A7M.
We used the same dielectric constants and salt concentration as in previous grid-convergence studies.
Other simulation parameters are listed in Table \ref{tab:sim_params_performance}.

\begin{table}[]
    \centering
    \begin{tabular}{lc}
    \hline
    \gmres tolerance          & $10^{-4}$ \\
    \# regular quadrature points  & 6    \\
    \fmm expansion order      & 5   \\
    \fmm $\ncrit$             & 500  \\
    \hline
    \end{tabular}
    \caption{Simulation parameters used in the performance study for 1A7M.}
    \label{tab:sim_params_performance}
\end{table}

To cover a wide range of problem sizes, we used five surface discretizations, with the number of elements ranging from 8 thousand to 2 million.
Table \ref{tab:1A7M_time} presents the assembly time, the solution time and the number of iterations to converge in each case for both formulations.
The assembly time includes the time to pre-compute the \fmm invariant matrices, create sparse and singular assemblers and calculate preconditioners.
The algebraic convergence shows that the condition number grows as the problem size increases with the direct formulation, while it remains at the same level with the derivative formulation.

\begin{table*}[]
    \centering
    \begin{tabular}{c|cccc|cccc}
                                                                 & \multicolumn{4}{c|}{direct}                                                                                                                                                                       & \multicolumn{4}{c}{derivative}                                                                                                                                                                        \\ \hline
    \begin{tabular}[c]{@{}c@{}}number of\\ elements\end{tabular} & \begin{tabular}[c]{@{}c@{}}total\\ time (s)\end{tabular} & \begin{tabular}[c]{@{}c@{}}assembly\\ time (s)\end{tabular} & \begin{tabular}[c]{@{}c@{}}GMRES\\ time (s)\end{tabular} & \# iterations & \begin{tabular}[c]{@{}c@{}}total\\ time (s)\end{tabular} & \begin{tabular}[c]{@{}c@{}}assembly\\ time (s)\end{tabular} & \begin{tabular}[c]{@{}c@{}}GMRES\\ time (s)\end{tabular} & \# iterations \\ \hline
    7884                                                         & 27.4                                                     & 7.6                                                         & 19.8                                                     & 33            & 31.2                                                     & 11.1                                                        & 20.1                                                     & 12            \\
    32996                                                        & 89.5                                                     & 12.5                                                        & 77.0                                                     & 50            & 72.7                                                     & 24.2                                                        & 48.5                                                     & 11            \\
    133512                                                       & 403.9                                                    & 34.6                                                        & 369.3                                                    & 92            & 197.3                                                    & 69.3                                                        & 128.0                                                    & 11            \\
    536220                                                       & 1570.7                                                   & 126.2                                                       & 1444.5                                                   & 120           & 646.7                                                    & 261.5                                                       & 385.2                                                    & 11            \\
    2148640                                                      & 7107.8                                                   & 500.7                                                       & 6607.1                                                   & 158           & 2424.3                                                   & 1064.4                                                      & 1359.9                                                   & 11           
    \end{tabular}
    \caption{Assembly and solution times of calculating the solvation energy of 1A7M, using the direct and derivative (exterior) formulations.
    6 regular quadrature points were used per element and the \fmm expansion order was set to 5.}
    \label{tab:1A7M_time}
\end{table*}

In our implementation, each iteration in direct formulation requires $8$ \fmm evaluations, whereas each iteration in the derivative formulation requires 19, making it more than twice as expensive.
That explains why the direct formulation leads to a shorter solution time (\gmres time), despite a slower convergence, in the smallest case.
For larger problem sizes, faster convergence in the derivative formulation offsets the larger cost per iteration.

\begin{figure*}[t]
\centering
    \subfloat[][Assembly time]{\includegraphics[width=0.33\textwidth]{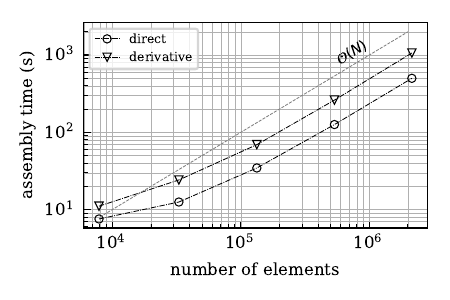}
        \label{fig:1A7M_assembly_time}}
   \subfloat[][Average evaluation time.]{\includegraphics[width=0.33\textwidth]{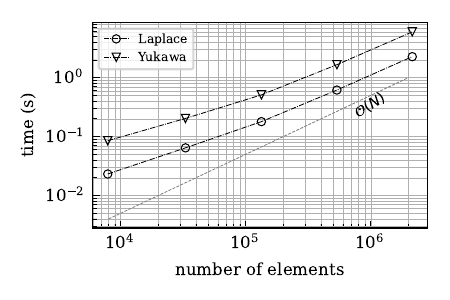}
        \label{fig:1A7M_fmm}}
    \subfloat[][Overall memory consumption in GB.]{\includegraphics[width=0.33\textwidth]{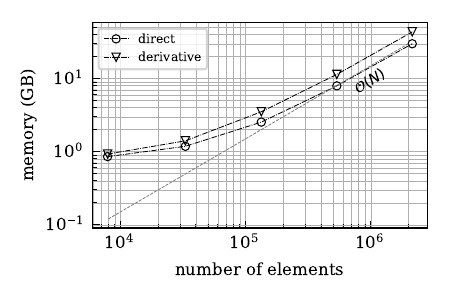}
        \label{fig:1A7M_memory}}\\
    \subfloat[][]{\includegraphics[width=0.4\textwidth]{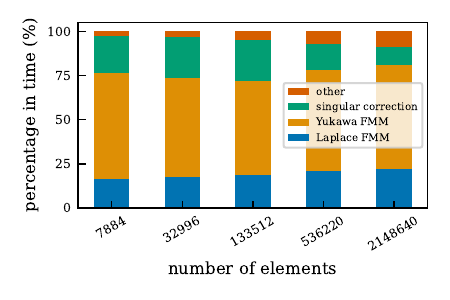}
        \label{fig:1A7M_gmres_direct}}
  \subfloat[][]{\includegraphics[width=0.4\textwidth]{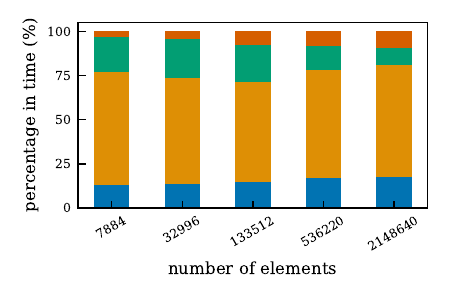}
        \label{fig:1A7M_gmres_derivative}}
    \caption{Performance on 1A7M;
    6 regular quadrature points per element; \fmm expansion order set to 5 to achieve 5 digits of accuracy. Problem size represented by number of elements, $N$. Evaluation time (b) is an average for 1 Laplace \fmm evaluation and 1 Yukawa \fmm evaluation across all iterations in GMRES using direct formulation.
    \textbf{d},\textbf{e}, Time breakdown of \gmres in percentage using direct formulation (\textbf{d}) and derivative formulation (\textbf{e}).}
\end{figure*}

As for the assembly time, the derivative formulation is about $2\times$ slower, since it needs to construct twice as many operators as the direct formulation.
In addition, the two hypersingular operators make it even more involved.
Figure \ref{fig:1A7M_assembly_time} shows the linear scaling of the assembly time with respect to $N$.

Next, we confirm that the time complexity of mat-vecs in \gmres is also $\mathcal{O}(N)$.
The Poisson equation requires \fmm with a Laplace kernel and the linearized Poisson-Boltzmann equation requires \fmm with a Yukawa kernel.
As mentioned, each iteration involves multiple \fmm evaluations: 4 Laplace {\fmm}s and 4 Yukawa {\fmm}s for the direct formulation, 8 and 11 for the derivative formulation.
We averaged the time spent on 1 Laplace \fmm and 1 Yukawa \fmm respectively using direct formulation, and plotted it with respect to $N$ in Figure \ref{fig:1A7M_fmm}.
Using an \fmm order of 5, we achieve about 5 digits of accuracy in each mat-vec.
The timings and linear scaling substantiate the efficiency of our \fmm implementation.
In the largest case with over 12 million quadrature points, one Laplace \fmm costs $2.3$s and one Yukawa \fmm costs $6.0$s to compute.

In Bempp, the matrix-vector product has the shape $A\mathbf{x} = P_2^T (G - C)P_1 \mathbf{x} + S \mathbf{x}$ (see Equation \ref{eq:bempp_fmm_matvec}), where the dominant costs are the \fmm evaluation of the Green's function matrix $G$ and the on-the-fly evaluation of the singular correction matrix $C$. Moreover, as stated above for the full $2\times 2$ block system a number of \fmm passes together with corresponding singular corrections need to be performed.
Therefore, the \gmres time reported here consists of \fmm time, singular correction time, as well as the time spent on other steps in the \gmres algorithm.
Figure \ref{fig:1A7M_gmres_direct} and \ref{fig:1A7M_gmres_derivative} show the time breakdown of \gmres in percentages.
In all cases, \fmm evaluations dominate the solution time.

We also measured the peak memory usage using the GNU time command \texttt{/usr/bin/time -v}. (Some Linux distributions do not ship with GNU time.)
We observed a linear space complexity as shown in Figure \ref{fig:1A7M_memory}.
The largest case, with more than $2$ million elements, requires $30$GB for the direct formulation and $44$GB for the derivative formulation.

\paragraph{Performance comparison with APBS} \label{result_comparison_apbs}
In this section, we compare the performance of Bempp and the finite-difference solver in APBS when computing the solvation energy of 1RCX.
Comparing the performance of two solvers is only fair at matching accuracy.
This is because the user has several runtime choices to make that affect both runtime and accuracy.
In the absence of an analytical solution, it is necessary to complete a grid convergence study with each solver to obtain an extrapolated solution at infinitely refined mesh, using Richardson extrapolation, based on which we can compute the errors.

We parameterized the molecule with \texttt{pdb2pqr} and the \texttt{charmm} force field. 
For APBS, we chose the \texttt{mg-auto} solver and used \texttt{dime} to specify the grid size.
Using three meshes with a constant refinement ratio of $1.34$, the observed order of convergence is $\mathcal{O}(h^{1.45})$ with respect to the grid spacing $h$.
Mesh sizes and the results are summarized in Table \ref{tab:1RCX_apbs}.
The solvation energy computed from the finest mesh is $-10803.07$ kcal/mol.
The memory cost grows linearly and the time to solution grows quadratically with respect to the grid size.

The surface meshes for Bempp have the element density set to 2, 4, 8 and 16 elements per $\si{\angstrom}^{2}$, respectively.
The performance of Bempp improves, without affecting the final accuracy, using 3 quadrature points for regular integrals and \fmm\ order equal to 3.
With the derivative formulation, the observed order of convergence is $\mathcal{O}(N^{-1.4})$, equivalent to $\mathcal{O}(h^{2.8})$.
The detailed results are listed in Table \ref{tab:1RCX_bempp}.
The solvation energy computed from the finest mesh is $-10997.16$ kcal/mol.

\begin{figure*}
    \begin{center}
        \includegraphics[width=0.4 \textwidth]{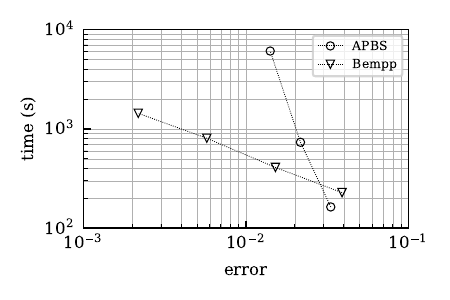}
        \label{fig:time_performance_comparsion}\qquad
        \includegraphics[width=0.4 \textwidth]{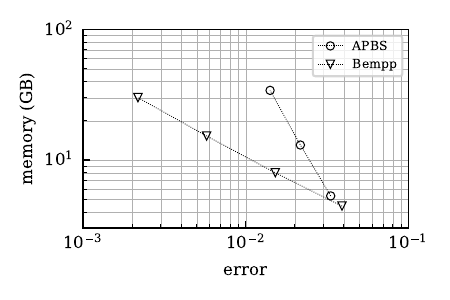}
        \label{fig:memory_performance_comparison}
    \end{center}
    \caption{Time and memory cost with respect to error for APBS and Bempp using 1RCX.
    Errors are measured against the extrapolated solution.
    }
    \label{fig:performance_comparison_apbs}
\end{figure*}

Using the errors from Table \ref{tab:1RCX_apbs} and \ref{tab:1RCX_bempp}, we plotted the time and space complexity with respect to accuracy obtained with Bempp and APBS, as shown in Figure \ref{fig:performance_comparison_apbs}.
Bempp displays better scaling in both time and space.
For this specific protein, the crossover point occurs around the 3\%-error mark.
Applications that require higher accuracy and involve larger structures will favor our boundary element solver.

\begin{table*}[]
    \centering
    \begin{tabular}{cc|cc|cc}
    grid size                   & h ($\si{\angstrom}$) & $\Delta G_{\mathrm{solv}}$ (kcal/mol) & error               & time (s) & memory (GB) \\ \hline
    $289 \times 289 \times 257$ & 0.776                & -11006.92                             & $3.3\times 10^{-2}$ & 164      & 5.3         \\
    $385 \times 385 \times 353$ & 0.582                & -10883.47                             & $2.2\times 10^{-2}$ & 734      & 13.0        \\
    $513 \times 513 \times 481$ & 0.437                & -10803.07                             & $1.4\times 10^{-2}$ & 6075     & 34.2       
    \end{tabular}
    \caption{Results from computing the solvation energy of 1RCX using the \texttt{mg-auto} solver in APBS.
    Error is calculated based on the extrapolated solution.}
    \label{tab:1RCX_apbs}
\end{table*}

\begin{table*}[]
    \centering
    \begin{tabular}{cc|cc|cc}
    N       & element density ($1/ \si{\angstrom}^2$) & $\Delta G_{\mathrm{solv}}$ (kcal/mol) & error               & time (s) & memory (GB) \\ \hline
    237416  & 2                                       & -11401.11                             & $3.9\times 10^{-2}$ & 229      & 4.4         \\
    483796  & 4                                       & -11139.84                             & $1.5\times 10^{-2}$ & 410      & 8.0         \\
    962664  & 8                                       & -11036.34                             & $5.7\times 10^{-3}$ & 806      & 15.3        \\
    1943736 & 16                                      & -10997.16                             & $2.2\times 10^{-3}$ & 1435     & 30.1       
    \end{tabular}
    \caption{Results from computing the solvation energy of 1RCX using Bempp-Exafmm and derivative formulation.
    3 quadrature points for regular integrals and \fmm\ order set to 3.
    Error is calculated based on the extrapolated solution.}
    \label{tab:1RCX_bempp}
\end{table*}

\paragraph{Solvation energy of a Zika virus} \label{result_zika}

\begin{figure*}
    \subfloat[][]{\includegraphics[width=0.25\textwidth]{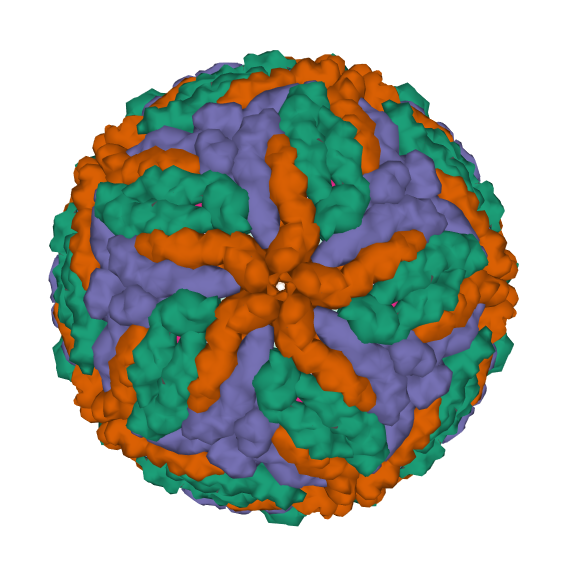}
        \label{fig:6CO8_assembly}}
\subfloat[][]{\includegraphics[width=0.75\textwidth]{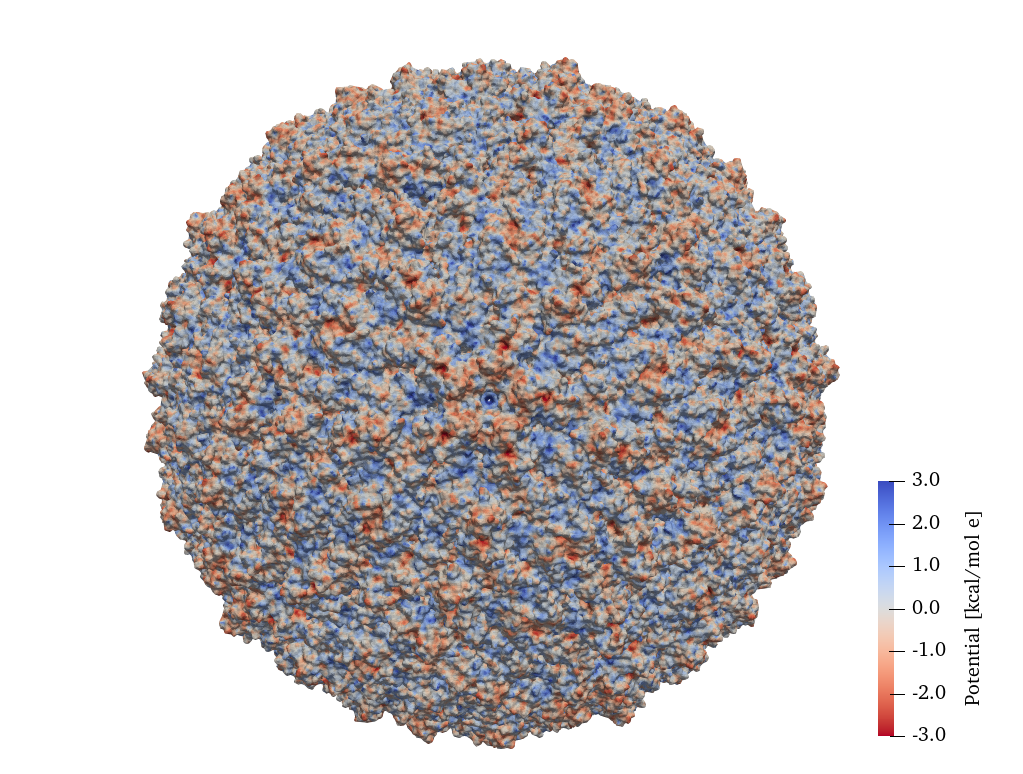}
        \label{fig:6CO8_potential}}
    \caption{\textbf{a}, Structure of Zika virus (PDB code 6CO8) in assembly view. Each color indicates a polymer chain.
    \textbf{b}, Surface electrostatic potential of a Zika virus.
    The color bar is in units of [kcal/mol$\cdot$e].
    Visualization generated using ParaView.
    The starfish pattern seen in the polymer-chain colorization of Figure \ref{fig:6CO8_assembly} is faintly visible in the potential.
    }
\end{figure*}

Finally, we present a more challenging problem that studies the solvation energy of the Zika virus (PDB code 6CO8), whose structure \cite{sevvana2018refinement} is shown in Figure \ref{fig:6CO8_assembly}.
We parameterized the molecular structure with the \texttt{amber} \cite{caseAmberBiomolecularSimulation2005} force field, and generated a mesh on the SES using \texttt{Nanoshaper}. (Mesh generation took less than 5 min.)
The prepared structure contains about 1.6 million atoms and our mesh has around 10 million boundary elements and a total surface area of $3.14\times 10^6 {\si{\angstrom}}^{2}$, corresponding to an element density of $\qty{3.16}{\angstrom}^{-2}$.
In this experiment, 3 quadrature points were used for regular Galerkin integrals over disjoint elements.
The \fmm expansion order was set to 4 obtain 4 digits of accuracy in mat-vecs and the tolerance of \gmres was $10^{-4}$.

\begin{table*}[]
    \centering
    \begin{tabular}{c|c|ccccc}
                 & \begin{tabular}[c]{@{}c@{}}$\Delta G_{\mathrm{solv}}$\\ (kcal/mol)\end{tabular} & \begin{tabular}[c]{@{}c@{}}total \\ time (s)\end{tabular} & \begin{tabular}[c]{@{}c@{}}assembly \\ time (s)\end{tabular} & \begin{tabular}[c]{@{}c@{}}GMRES \\ time (s)\end{tabular} & \begin{tabular}[c]{@{}l@{}}memory\\ (GB)\end{tabular} & \# iterations \\ \hline
    direct     & -116587.5                                               & 11005.4                                                   & 1534.5                                                       & 9470.9                                                    & 109.7                                                 & 105           \\
    derivative & -116254.9                                               & 8370.3                                                    & 3553.9                                                       & 4816.4                                                    & 152.0                                                 & 18            
    \end{tabular}
    \caption{Results of computing the solvation energy of a Zika virus with Bempp using both the direct and derivative formulations.
    3 regular quadrature points were used per element and the \fmm expansion order was set to 4.
    The mesh generation time is minor and not included in the total time.
    To verify our result, we ran the same case with \pygbe using the direct formulation on a workstation with a 24-core CPU.
    The solvation energy computed from \pygbe is -117261.1 [kcal/mol].
    }
    \label{tab:6CO8_result}
\end{table*}

Table \ref{tab:6CO8_result} summarizes the results and performance.
Again, we confirmed that the derivative formulation yields a well-conditioned system, which converged in 18 iterations and took less than 1.5 hours to solve.
By contrast, the direct formulation took almost twice as long to converge.

In this case, we also verified against \pygbe \cite{cooper2016pygbe}, a Python \bem library for biomolecular electrostatics.
Based on the solvation energy computed from \pygbe: $-117261.1$ [kcal/mol], the relative difference of our result is 0.6\% with the direct formulation and 0.9\% with the derivative formulation.
Figure \ref{fig:6CO8_potential} shows the computed electrostatic potential on the surface.

\paragraph{Reproducibility package}

Besides releasing all our software with an open-source license, we spared no effort to maximize the reproducibility of this work, compiling a ``repro-pack" in a version-control repository.
It contains all raw data from the experiments and a small Python package---\texttt{bempp\_pbs}, to facilitate running PB simulations with Bempp.
\texttt{bempp\_pbs} comprises a collection of scripts, including all post-processing scripts that are necessary to produce every result presented in this section, and driver and utility scripts for different formulations, with which readers can run these cases on their own hardware.
In addition, the ``repro-pack" also provides a Jupyter notebook for each study to generate secondary data and results.
The notebook corresponding to section \ref{result_conditioning} showcases how we could run a simple case of a spherical molecule using two formulations with just a few lines of code, and then compare their conditioning quantitatively with the help of other scientific Python tools.
These supplementary materials are included in our paper's GitHub repository at \href{https://github.com/barbagroup/bempp\_exafmm\_paper/}{https://github.com/barbagroup/bempp\_exafmm\_paper/}, which is also archived on Zenodo at \href{http://doi.org/10.5281/zenodo.4568951}{doi:10.5281/zenodo.4568951}.
We made separate archival deposits of input data (meshes and \texttt{pqr} files ) on the Zenodo service.
The deposit can be downloaded from \href{http://doi.org/10.5281/zenodo.4568768}{doi:10.5281/zenodo.4568768}.

\section{Discussion} \label{sec:discussion}
With this paper, we introduce a new platform for computational investigations in biomolecular electrostatics, combining high-performance with high researcher productivity. 
Bempp-Exafmm integrates one of the most trusted boundary element software packages with one of the most performant fast-summation libraries using multipole methods. 
A Python entry point gives researchers ease of use, while enabling computational research at virus scale on standard workstations.
The software is open source under permissive public licenses, and developed in the open.

We present several results that confirm the usefulness of the platform, verify solution correctness with classic benchmarks, and showcase the performance. 
In section \ref{result_conditioning}, we compared the conditioning of the interior and exterior derivative formulations.
Despite the fact that both yield a well-conditioned system, where the condition number does not grow with the problem size, the exterior formulation always converges faster due to the clustering of its eigenvalues.
It shows a greater advantage over the interior formulation as the difference between $\epsilon_1$ and $\epsilon_2$ becomes larger.
Previous publications have used one or the other of these solution methods, but we found no comparison of the two in the literature.
Our experiments showed that the exterior formulation converges twice as fast with typical values of the permittivities: an important advantage.
Seeing this, we quickly computed condition numbers, made heatmaps of the matrices, and plotted eigenvalues on the complex plane---all interactively, in a Jupyter notebook---leading to an explanation of the different algebraic behaviors.
(Readers can find a final Jupyter notebook in the GitHub repository for this manuscript.)
This study also serves as example of how users can benefit from our high-productivity platform.
Through interactive computing, users can adapt various formulations, try out different problem setups, analyze intermediate results on-the-fly without the hassle of recompilation.

We performed two mesh-refinement studies to verify Bempp-Exafmm, using a spherical molecule with an off-center charge, and using a real biomolecule (bovine pancreatic trypsin inhibitor).
In the former study, we compared with the analytical solution; in the latter, we compared with an approximate value from Richardson extrapolation.
In both cases, we used the direct formulation and the exterior derivative formulation.
To reveal the discretization error, we set the \fmm expansion order to 10 to achieve 9 digits of accuracy.
The error of the computed solvation energy decays linearly with respect to $N$, as shown in Figure \ref{fig:sphere_convergence} and \ref{fig:5PTI_convergence}.
Our convergence result is consistent with those from \pygbe \cite{CooperBardhanBarba2014} and TABI \cite{GengKrasny2013}, where linear convergence is also observed with real molecules.

Binding energy calculations involve combining multiple solvation energies and thus are more challenging.
We computed the binding energy of 9 different complexes and compared our finest-grid results with results from 5 grid-based PB solvers using a grid spacing of \qty{0.2}{\angstrom}.
Figure \ref{fig:bind} displays that our results fall within the range of those from other solvers, which further substantiates the correctness of Bempp-Exafmm.

In section \ref{result_performance}, we elaborate on the performance of Bempp-Exafmm for different problem sizes using both formulations.
The linear complexity of the assembly time (Figure \ref{fig:1A7M_assembly_time}) and \fmm time (Figure \ref{fig:1A7M_fmm}) guarantees the overall linear time complexity of Bempp-Exafmm, which, together with the linear space complexity shown in Figure \ref{fig:1A7M_memory}, makes it feasible to perform large-scale simulations on a workstation.
Table \ref{tab:1A7M_time} lists the timings in detail.
Despite being ill-conditioned, the direct formulation still shows an advantage in terms of the overall time for smaller problem sizes.
Conversely, the derivative formulation shines in larger problems.
The crossover point should be problem- and hardware-specific.

We compared the performance of Bempp and APBS across different levels of accuracy using a moderate-sized structure 1RCX.
We used three grid spacings for APBS: \qtylist{0.776;0.582;0.437}{\angstrom}, and four mesh densities for Bempp: 2, 4, 8, 16 elements per $\si{\angstrom}^{2}$ in our test.
For each code, the errors are computed based on the corresponding extrapolated solution.
Though APBS has an edge over Bempp for low accuracy calculations (error $>3\%$), Bempp performs better when higher accuracy is required due to better scaling.
We want to admit that APBS, as well as many other grid-based PB solvers, may run faster than Bempp for smaller proteins, as the crossover point in Figure \ref{fig:performance_comparison_apbs} will shift left.
However, most of these solvers are not suitable to handle large structures at a decent accuracy on a workstation due to the excessive time and memory cost, as suggested by the scaling in Figure \ref{fig:performance_comparison_apbs}.

Finally, we computed the solvation energy of a Zika virus using both formulations and verified our results against \pygbe in \ref{result_zika}.
The linear system, for a mesh containing 10 million boundary elements, was solved in 80 minutes on a single node.
It shows that the performance of our code is in the same ballpark as other state-of-the-art fast BEM PB solvers \cite{GengKrasny2013,ZhangPengHuangPitsianisSunLu2015,CooperBardhanBarba2014} and gives us confidence in its capability of solving virus-scale problems.

Poisson-Boltzmann solvers have been around for decades, available as stand-alone applications and web servers, and using a variety of solution approaches \cite{JurrusETal2018} ranging from finite difference, to finite element, and boundary element methods.
Some solvers are integrated into a number of computational workflows that use them for mean field potential visualization \cite{HumphreyETal1996} and free energy calculations \cite{MillerETal2012,KumariETal2014}, usually interfaced through bash or Python scripts.
Our approach also becomes useful as part of existing computational workflows, such as MMPBSA methods \cite{WangETal2018}.
The Poisson-Boltzmann model has known limitations, being the rigid approximation of the solute a specially difficult one at large scales. 
Regardless, the electrostatic potential and polar solvation free energy are valuable information even at large scales \cite{LiETal2019}, for analysis of, for example, the size and structure of a virus \cite{SiberETal2012}, or its pH dependence \cite{RoshalETal2019}. 

The modern design of Bempp is built such that high-performance computations are accessible from a high-productivity language.
This makes our effort stand out in the current landscape of Poisson-Boltzmann solvers in three ways: interoperability, ease of use, and robustness. 
\begin{enumerate}
\item Interoperability: Bempp is written in Python, and hence, is callable from a Jupyter notebook. This fits naturally in any computational workflow that uses Jupyter notebooks, for example, with openMM \cite{EastmanETal2017}, Biobb \cite{AndrioETal2019}, MDAnalysis \cite{GowersETal2019}, pytraj \cite{RoeCheatham2013}, or PyMOL \cite{PyMOL}. The Jupyter Notebook becomes a computational glue across models and scales; no interface script required. 

\item Ease of use: Python and Jupyter notebooks are widely used, even in non-computational settings. Bempp is easily installed through \texttt{conda}, and gives a result in less than 20 lines of code. This, moreover, using parallel and state-of-the-art algorithms in a way that is almost transparent to the user, allowing for large-scale simulations on workstations or small clusters.
A thin layer separates the application and Bempp, giving a more experienced user access to develop new models, for example, through the \fmm-\bem coupling capability of Bempp.

\item Robustness: Bempp is actively developed with high standards of software engineering, such as unit and system testing, continuous integration, etc. It was originally designed for scattering problems, impacting a large group of people, well beyond the molecular simulation community. This builds high trust and reliability of the code, as it is thoroughly tested in a diverse set of applications. The software has a better chance to survive in the long term, and any improvements done by people in other domains will have an effect on its use to solve the Poisson-Boltzmann equation. 

\end{enumerate}

Many popular molecular simulation software packages exist, designed for different applications, scales, quantities of interest, etc.
This has led to community-wide efforts, such as BioExcel (\url{https://bioexcel.eu/}) and MolSSI (\url{http://molssi.org}), that are looking for a common ground between them, as well as promoting good software development practices for robust and easy-to-use codes.
This standard is very much aligned with our work.

Modern research software efforts today aim for the union of high performance and high researcher productivity.
A vigorous trend is unmistakable towards empowering users with interactive computing, particularly using Jupyter notebooks. 
Our work contributes a platform for interactive investigations in biomolecular electrostatics that is easy to use, easy to install, highly performant, extensible, open source and free.
We contemplate a bright future for science domains that gel community efforts to jointly develop and curate software tools with similar philosophy.

\small{
\section{Methods}\label{sec:methods}
\paragraph{Boundary integral formulation of electrostatics in molecular solvation}\label{s:formulation}

Biomolecules in an ionic solvent or water can be represented with a model where the solvent is a continuous dielectric: the so-called implicit-solvent model.
The molecule (solute) is a dielectric cavity with partial charges at the atomic locations,  represented as a collection of Dirac delta functions.
In  the sketch of Figure \ref{fig:implicit_solvent}, the molecular cavity is region $\Omega_1$, the infinite medium  is $\Omega_2$, and the point charges are $q_k$.
The interface $\Gamma$ represents the molecular surface and can be determined through several approaches: van der Waals radii, Gaussian surface, solvent-accessible surface, and solvent-excluded surface. 
We use the latter.
A solvent-excluded surface is built by tracking the contact points of a (virtual) spherical probe of  $\sim$ 1.4 \AA\ in radius, the size of a water molecule, as it rolls around the solute's atoms (with their corresponding van der Waals radii). 
In this setup, we can compute the change in electrostatic potential as the interior region is charged up with the solute's partial charges.
The solvent usually consists of water (dielectric constant $\epsilon_2\approx80$) with salt ions that are free to move around, forced by the electric field. 
At equilibrium, the salt ions are in a Boltzmann distribution, leading to the linearized Poisson-Boltzmann equation for the potential in $\Omega_2$, considering a screening factor known as the inverse of the Debye length ($\kappa$). 
The electrostatic potential in the solute cavity follows Poisson's equation in a low dielectric medium ($\epsilon_1\approx2\textrm{--}4$), with the solute's point charges as sources.
With interface conditions on the molecular surface $\Gamma$, where the potential and electric displacement must be continuous, this results in the following system of equations:
\begin{align} \label{eq:pde}
\nabla^2\phi_1 &= \frac{1}{\epsilon_1}\sum_k q_k\delta(\mathbf{r},\mathbf{r}_k) \text{ in the solute ($\Omega_1$),}\nonumber\\
(\nabla^2-\kappa^2)\phi_2 &= 0 \text{ in the solvent ($\Omega_2$),}\nonumber\\
\phi_1 &= \phi_2 \quad \epsilon_1\frac{\partial \phi_1}{\partial\mathbf{n}} = \epsilon_2\frac{\partial \phi_2}{\partial\mathbf{n}} \text{ on the interface ($\Gamma$)}.
\end{align}
\begin{figure}
\centering
\includegraphics[width=0.2\textwidth]{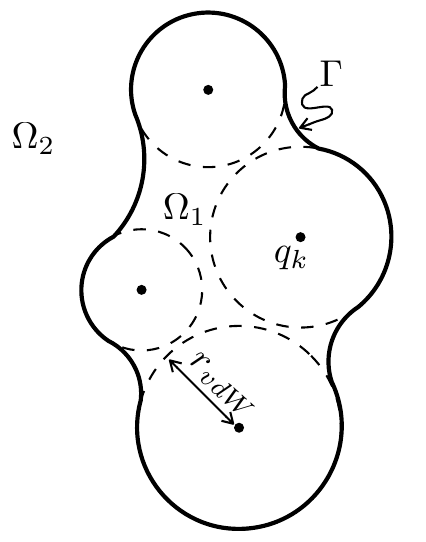}
\caption{Representation of a dissolved molecule with the implicit-solvent model. The solute ($\Omega_1$) and solvent ($\Omega_2$) regions are interfaced by the solvent-excluded surface ($\Gamma$), and $q_k$ and $r_{vdW}$ are the atomic charge and radii, respectively.}
\label{fig:implicit_solvent}
\end{figure}

Equation \eqref{eq:pde} can be re-written as an integral equation on $\Gamma$ via Green's second identity, yielding:
\begin{align} \label{eq:volume_potential}
\phi_{1}+ K_{L}^{\Omega_1}(\phi_{1,\Gamma}) -  V_{L}^{\Omega_1} \left(\frac{\partial}{\partial \mathbf{n}}  \phi_{1,\Gamma}  \right) & = \frac{1}{\epsilon_1} \sum_{k=0}^{N_q}  \frac{q_k}{4\pi|\mathbf{r}_{\Omega_1} - \mathbf{r}_k|}  \quad \text{on $\Omega_1$,} \nonumber \\
\phi_{2} - K_{Y}^{\Omega_2}(\phi_{2,\Gamma}) + V_{Y}^{\Omega_2} \left( \frac{\partial}{\partial \mathbf{n}} \phi_{2,\Gamma} \right) & = 0 \quad \text{on $\Omega_2$,}
\end{align}
where $\phi_{1,\Gamma} = \phi_1(\mathbf{r}_\Gamma)$ and $\phi_{2,\Gamma} = \phi_2(\mathbf{r}_\Gamma)$ are evaluated on $\Gamma$ approaching from $\Omega_1$ and $\Omega_2$, respectively. $K$ and $V$ are the double- and single-layer potentials for the Laplace (subscript $L$) and Yukawa (subscript $Y$) kernels, defined as:
\begin{align}\label{eq:single_double}
V^\Omega_{L,Y}(\varphi) = \oint_\Gamma g_{L,Y}(\mathbf{r}_\Omega,\mathbf{r}')\varphi(\mathbf{r}')d\mathbf{r}'\nonumber\\
K^\Omega_{L,Y}(\varphi) = \oint_\Gamma \frac{\partial g_{L,Y}}{\partial\mathbf{n}'}(\mathbf{r}_\Omega,\mathbf{r}')\varphi(\mathbf{r}')d\mathbf{r}',\nonumber\\
\end{align}
where $\varphi(\mathbf{r})$ is a distribution over $\Gamma$, and $g_L(\mathbf{r},\mathbf{r}')=\frac{1}{4\pi|\mathbf{r}-\mathbf{r}'|}$ and $g_Y(\mathbf{r},\mathbf{r}')=\frac{e^{-\kappa|\mathbf{r}-\mathbf{r}'|}}{4\pi|\mathbf{r}-\mathbf{r}'|}$ are the free-space Green's function of the Laplace and linearized Poisson-Boltzmann equations, respectively. 

We can use Equation \eqref{eq:volume_potential} to compute $\phi_\Gamma$ and $\partial\phi_\Gamma/\partial\mathbf{n}$ with either the \emph{direct}~\cite{YoonLenhoff1990} or \emph{derivative}~\cite{JufferETal1991} (also known as \emph{Juffer}) formulations.
The simpler direct formulation results from evaluating $\phi_1$ and $\phi_2$ in the limit as $\mathbf{r}$ approaches $\Gamma$, and applying the interface conditions from Equation \eqref{eq:pde}, giving:
\begin{align} \label{eq:direct}
\frac{\phi_{1,\Gamma}}{2}+ K_{L}^{\Gamma}(\phi_{1,\Gamma}) -  V_{L}^{\Gamma} \left(\frac{\partial}{\partial \mathbf{n}}  \phi_{1,\Gamma}  \right) & = \frac{1}{\epsilon_1} \sum_{k=0}^{N_q}  \frac{q_k}{4\pi|\mathbf{r}_{\Gamma} - \mathbf{r}_k|} \nonumber \\
\frac{\phi_{1,\Gamma}}{2} - K_{Y}^{\Gamma}(\phi_{1,\Gamma}) + \frac{\epsilon_1}{\epsilon_2}V_{Y}^{\Gamma} \left( \frac{\partial}{\partial \mathbf{n}} \phi_{1,\Gamma} \right) & = 0
\end{align}
This formulation is ill-conditioned, since the condition number of the resulting matrix grows unbounded with the number of discretization elements. 
An alternative with better conditioning was derived by Juffer \emph{et al.} \cite{JufferETal1991} by taking the normal derivative of Equation \eqref{eq:volume_potential}, and coupling both $\phi$ and $\partial\phi/\partial\mathbf{n}$ on the boundary as follows:

\begin{align}\label{eq:juffer}
    &\begin{multlined}[t][0.48\textwidth] \frac{\phi_{1,\Gamma}}{2}\left(1+\frac{\epsilon_2}{\epsilon_1}\right) - \left(\frac{\epsilon_2}{\epsilon_1}K_Y^\Gamma - K_L^\Gamma\right)(\phi_{1,\Gamma}) \\
    + \left(V_Y^\Gamma - V_L^\Gamma\right)\left( \frac{\partial}{\partial \mathbf{n}} \phi_{1,\Gamma} \right) = \sum_{k=0}^{N_q}  \frac{q_k}{4\pi\epsilon_1|\mathbf{r}_{\Gamma} - \mathbf{r}_k|}
    \end{multlined} \nonumber \\
    &\begin{multlined}[t][0.48\textwidth] \left(W_Y^\Gamma - W_L^\Gamma\right)(\phi_{1,\Gamma}) +  \frac{1}{2}\frac{\partial\phi_{1,\Gamma}}{\partial\mathbf{n}}\left(1+\frac{\epsilon_1}{\epsilon_2}\right) \\
    + \left(\frac{\epsilon_1}{\epsilon_2}K_Y^{\prime\Gamma} - K_L^{\prime\Gamma}\right)\left( \frac{\partial}{\partial \mathbf{n}} \phi_{1,\Gamma} \right) = \sum_{k=0}^{N_q}  \frac{\partial}{\partial\mathbf{n}_\mathbf{r}}\left(\frac{q_k}{4\pi\epsilon_1|\mathbf{r}_{\Gamma} - \mathbf{r}_k|}\right)
    \end{multlined}
\end{align}
Here, we use the adjoint double-layer ($K'$) and hypersingular ($W$) operators, which are defined as
\begin{align}\label{eq:adj_hyp}
K^{\prime\Gamma}_{L,Y}(\varphi) = \oint_\Gamma \frac{g_{L,Y}}{\partial\mathbf{n}}(\mathbf{r}_\Gamma,\mathbf{r}')\varphi(\mathbf{r}')d\mathbf{r}'\nonumber\\
W^\Gamma_{L,Y}(\varphi) = - \oint_\Gamma \frac{\partial^2 g_{L,Y}}{\partial\mathbf{n}'\partial\mathbf{n}}(\mathbf{r}_\Gamma,\mathbf{r}')\varphi(\mathbf{r}')d\mathbf{r}'\nonumber\\
\end{align}
A slightly modified version of Equation \eqref{eq:juffer} is used in the work from Lu and coworkers~\cite{LuETal2006,LuETal2009,ZhangETal2019}, where they scale the expressions by $\epsilon_1/\epsilon_2$, and solve for the exterior field. This gives
\begin{align}\label{eq:lu}
    &\begin{multlined}[t][0.48\textwidth] \frac{\phi_{2,\Gamma}}{2}\left(\frac{\epsilon_1}{\epsilon_2}+1\right) - \left(K_Y^\Gamma - \frac{\epsilon_1}{\epsilon_2}K_L^\Gamma\right)(\phi_{2,\Gamma}) \\
    + \left(V_Y^\Gamma - V_L^\Gamma\right)\left( \frac{\partial}{\partial \mathbf{n}} \phi_{2,\Gamma} \right) = \sum_{k=0}^{N_q}  \frac{q_k}{4\pi\epsilon_2|\mathbf{r}_{\Gamma} - \mathbf{r}_k|}
    \end{multlined} \nonumber \\
    &\begin{multlined}[t][0.48\textwidth] \frac{\epsilon_1}{\epsilon_2}\left(W_Y^\Gamma - W_L^\Gamma\right)(\phi_{2,\Gamma}) +  \frac{1}{2}\frac{\phi_{2,\Gamma}}{\partial\mathbf{n}}\left(1+\frac{\epsilon_1}{\epsilon_2}\right) \\
    + \left(\frac{\epsilon_1}{\epsilon_2}K_Y^{\prime\Gamma} - K_L^{\prime\Gamma}\right)\left( \frac{\partial}{\partial \mathbf{n}} \phi_{2,\Gamma} \right) = \sum_{k=0}^{N_q}  \frac{\partial}{\partial\mathbf{n}_\mathbf{r}}\left(\frac{q_k}{4\pi\epsilon_2|\mathbf{r}_{\Gamma} - \mathbf{r}_k|}\right)
    \end{multlined}
\end{align}

As we charge up the cavity, the solvent ions rearrange and polarize.
The resulting electrostatic potential is called a \emph{reaction} potential ($\phi_{reac}$), and we can write the following decomposition in $\Omega_1$ :
\begin{equation}
\phi_1 = \phi_{reac} + \phi_{coul},
\end{equation}
where $\phi_{coul}$ is the Coulombic potential from the solute point charges only.
Having $\phi_{1,\Gamma}$ and $\partial\phi_{1,\Gamma}/\partial\mathbf{n}$ from Equation \eqref{eq:direct} or Equation \eqref{eq:juffer}, we can compute $\phi_{reac}$ by subtracting out the Coulombic contribution in the right-hand side of Equation \eqref{eq:volume_potential}:
\begin{equation}\label{eq:phi_reac}
\phi_{reac} = -K_{L}^{\Omega_1}(\phi_{1,\Gamma}) +  V_{L}^{\Omega_1} \left(\frac{\partial}{\partial \mathbf{n}}  \phi_{1,\Gamma}  \right) 
\end{equation}

The thermodynamic work required to dissolve a molecule, known as solvation free energy, is usually divided into nonpolar and polar components.
The nonpolar part generates the empty solute-shaped cavity in the solvent, which is then charged by placing the partial charges inside the cavity, giving rise to a polar term in the energy. 
The work in charging is performed under $\phi_{reac}$, and it can be computed as:
\begin{equation} \label{eq:energy}
\Delta G^{polar}_{solv} = \frac{1}{2}\int_{\Omega_1} \rho\phi_{reac}d\mathbf{r} = \frac{1}{2}\sum_{k=1}^{N_q}q_k\phi_{reac}(\mathbf{r}_k).
\end{equation}

\paragraph{An introduction to Bempp}
Bempp \cite{Betcke2021} is a Python based boundary element library for the Galerkin discretization of boundary integral operators in electrostatics, acoustics and electromagnetics.
Bempp originally started as mixed Python/C++ library.
Recently, Bempp underwent a complete redevelopment and the current version Bempp-cl is written completely in Python with OpenCL kernels for the low-level computational routines that are just-in-time compiled for the underlying architecture during runtime.
To understand Bempp consider the simple boundary integral equation
$$
\int_{\Gamma} g(\mathbf{r}, \mathbf{r'}) \phi(\mathbf{r'})ds(\mathbf{r'}) = f(\mathbf{r})
$$
where $\Gamma\subset\mathbb{R}^3$ is the surface of a bounded domain $\Omega\subset\mathbb{R}^3$, and $g(\mathbf{r}, \mathbf{r'}) = \frac{1}{4\pi|\mathbf{r}-\mathbf{r'}|}$ is the electrostatic Green's function.
A Galerkin discretization of this equation takes the form
$$
A\mathbf{x} = b
$$
with $A_{ij} = \int_{\Gamma}\Psi_i(\mathbf{r})\int_{\Gamma}g(\mathbf{r}, \mathbf{r'})\phi_j(\mathbf{r'})ds(\mathbf{r'})ds(\mathbf{r})$ and $b_i = \int_{\Gamma}\psi_i(\mathbf{r})f(\mathbf{r})ds(\mathbf{r})$.
Here, the functions $\Psi_j$ are a finite dimensional basis of $n$ test functions and the $\phi_j$ are a finite dimensional basis of $n$ trial functions with the Galerkin solution being defined as $\phi=\sum_{i}\mathbf{x}_j\phi_j$.
Typical choices for the test and trial functions are either piecewise constant functions or continuous, piecewise linear functions over a surface triangulation of $\Gamma$. 
By default, Bempp explicitly computes the matrix $A$ by applying quadrature rules to the arising integrals.
The singularity of the Green's function needs to be accounted for in the quadrature rules for integration over adjacent or identical test/trial triangles \cite{ERICHSEN1998215}.
For well separated triangles standard triangle Gauss rules can be used for the quadrature.
The memory and computational complexity of this discretization is $\mathcal{O}(n^2)$, which is practical for problems of size up to twenty or thirty thousand elements on a single workstation, depending on the available memory and number of CPU cores.
Bempp-cl evaluates the quadrature routines with highly optimized OpenCL kernels that make use of explicit AVX2/AVX-512 acceleration on CPUs.

\paragraph{FMM-accelerated evaluation of integral operators}
We can split up the action of the discretized integral operator $A$ onto a vector $\mathbf{x}$ in the following way.
\begin{equation}
\label{eq:bempp_fmm_matvec}
A\mathbf{x} = P_2^T (G - C)P_1 \mathbf{x} + S \mathbf{x}.
\end{equation}
The matrices $P_1$ and $P_2$ are sparse matrices that convert the action of trial and test functions onto weighted sums over the quadrature points.
The matrix $G$ is a large dense matrix that contains the Green's function evaluation $g(\mathbf{r}_i, \mathbf{r}_j')$ over all quadrature points $\mathbf{r}_i$ and $\mathbf{r}_j'$ across all triangles.
The matrix $C$ is a sparse correction matrix that subtracts out the Green's function values over quadrature points associated with  adjacent triangles.
This is done since these triangles require a singularity adapted quadrature rule.
By explicitly subtracting out these contributions through the matrix $C$, we can use any code for the fast evaluation of particle sums of the type appearing in the $G$ matrix without the requirement to communicate to the summation code the geometry and singularity structure induced by the triangles of the surface mesh, a functionality that most such codes do not offer in any case.
Finally, the matrix $S$ contains the contributions of $A$ arising from singularity adapted quadrature rules across adjacent or identical test/trial triangles.
This matrix is also highly sparse.

We explicitly compute the matrices $P_1$, $P_2$ and $S$, and keep them in memory using sparse storage.
The matrix $C$ is evaluated on the fly for each vector $\mathbf{x}$ through a fast OpenCL kernel.
This leaves the matrix $G$.
The action of $G$ on the vector $\mathbf{y}=P_1 \mathbf{x}$ can be considered as a black-box to evaluate sums of the form
\begin{align}\label{eq:nbody_sum}
s(\mathbf{x}_i) = \sum_j g(\mathbf{r}_i, \mathbf{r}_j')q_j.
\end{align}
To evaluate this sum we use the C++ Exafmm library, a highly performant library that implements the kernel-independent fast multipole method (\kifmm) to approximately evaluate sums of the above form.
The complexity of this evaluation is $\mathcal{O}(N)$, where $N$ is the product of the number of surface triangles and the number of regular quadrature points per triangle.
The linear complexity means that we can scale the evaluation of the discretized integral operator from tens of thousands to millions of elements, allowing us to solve large electrostatic simulations on a single workstation. Details of the \fmm implementation are discussed in the following section.

\paragraph{Fast multipole method}

The fast multipole method (\fmm) is an algorithm that can reduce the quadratic time and space complexity of such matrix-vector multiplication down to $\mathcal{O}(N)$.
In the context of \fmm, $\{\mathbf{r}_i\}$ and $\{\mathbf{r}_j'\}$ in Equation \ref{eq:nbody_sum} are often referred to as the set of targets and sources respectively, with $\{q_j\}$ representing the source densities (charges).
The goal of \fmm is to efficiently compute the potential at $N$ targets $\{s_i\}$ induced by all $N$ sources and the kernel function $g$.
Following the common notations in the literature, we use $\mathbf{x}_i$ and $\mathbf{y}_j$, instead of $\mathbf{r}_i$ and $\mathbf{r}_j'$, to denote targets and sources respectively in this subsection.

The \fmm algorithm builds upon two fundamental ideas: (1) approximating the far-range interactions between distant clusters of sources and targets using low-rank methods, while computing the near-range interactions exactly, and (2) partitioning the domain using a tree structure to maximize the far-range portion in the computation.

To construct the octree, we first create a cube that encloses all sources/targets and then recursively subdivide the domain until each cube at the finest level only contains a constant number of points.
Figure \ref{fig:near_far_decomp} depicts a 3-level quadtree.
The potentials of targets in node $B$ consist of three contributions: the influence from sources in the near-field of $B$: $\mathcal{N}(B)$, in the interaction list of $B$: $\mathcal{I}(B)$, and in the rest of the domain.
$B$'s near-field includes $B$ and its neighbors, where the interactions are computed exactly.
The remaining domain is in $\mathcal{F}(B)$, $B$'s far-field.
In $\mathcal{F}(B)$, the nodes that are the children of $B$'s parent's neighbors but are not adjacent to $B$ compose $\mathcal{I}(B)$, the interaction list of $B$, whose contributions to $B$ are approximated by low-rank methods.
The contributions from the rest of the far-field are approximated at coarser levels via $B$'s ancestors.

The classic \fmm \cite{greengard1987fast, cheng1999fast} relies on truncated analytical expansions to approximate far-field interactions, whereas its kernel-independent variant \cite{ying2004kernel} uses equivalent densities (charges) instead.
In \kifmm, each node is associated with upward and downward equivalent densities (see Figure \ref{fig:multipole} and \ref{fig:local}), the analog of multipole and local expansions in the analytical \fmm.
The upward equivalent densities $q^{B,u}$ are used to approximate the influence of sources in $B$ on targets in $\mathcal{F}(B)$;
the downward equivalent densities $q^{B,d}$ are used to approximate the influence of sources in $\mathcal{F}(B)$ on sources in $B$.
To find these densities, we match the potential of equivalent densities to the potential of actual sources at the check surfaces:
\begin{align}\label{eq:multipole_local}
    \sum_{\mathbf{y}_{j} \in B} g\left(\mathbf{x}_{i}^{B,u}, \mathbf{y}_{j}\right) q_{j} &= \sum_{j} g\left(\mathbf{x}_{i}^{B,u}, \mathbf{y}^{B,u}_{j}\right) q^{B,u}_{j}, \quad \forall i  \nonumber \\
    \sum_{\mathbf{y}_{j} \in \mathcal{F}(B)} g\left(\mathbf{x}_{i}^{B,d}, \mathbf{y}_{j}\right) q_{j} &= \sum_{j} g\left(\mathbf{x}_{i}^{B,d}, \mathbf{y}^{B,d}_{j}\right) q^{B,d}_{j}, \quad \forall i
\end{align}
We then solve the linear systems for $\{q^{B,u}_{j}\}$ and $\{q^{B,d}_{j}\}$.
Here, $\mathbf{x}_{i}^{B}$ and $\mathbf{y}_{j}^{B}$ denote the discretization points of the check surface and equivalent surface of $B$ respectively.

The algorithm also defines the following operators:
\begin{itemize}
    \item particle-to-multipole (P2M): For a leaf node $B$, compute $B$'s upward equivalent densities, \ie multipole expansion, from the sources in $B$. (Figure \ref{fig:multipole})
    \item multipole-to-multipole (M2M): For a non-leaf node $B$, evaluate $B$'s multipole expansion based on the multipole expansions of all $B$'s children. (Figure \ref{fig:translations} left)
    \item multipole-to-local (M2L): For a node $B$, evaluate $B$'s downward equivalent densities, \ie local expansion, by using the multipole expansions of all nodes in $\mathcal{I}(B)$. (Figure \ref{fig:translations} middle)
    \item local-to-local (L2L): For a non-leaf node $B$, add the contribution of $B$'s local expansion to the local expansions of $B$'s children. (Figure \ref{fig:translations} right)
    \item local-to-particle (L2P): For a leaf node $B$, evaluate $B$'s local expansion at the locations of targets in $B$.
    This step adds all far-field contribution to the potentials of targets in $B$. 
    \item particle-to-particle (P2P): For a leaf node $B$, evaluate the potential induced by all sources in $\mathcal{N}(B)$ directly.
\end{itemize}
As indicated by the arrows in Figure \ref{fig:translations}, translation operators in \kifmm share the same procedure: (1) evaluating the potentials on the check surface, and (2) solving the equation arising from matching the potentials for the equivalent densities.

Figure \ref{fig:fmm_sketch} outlines the complete \fmm algorithm.
During the upward pass, we compute P2M at all leaf nodes and perform M2M in post-order tree traversal.
Next, we compute M2L for all nodes.
Finally, we compute L2L in pre-order tree traversal, and perform L2P and P2P at all leaf nodes during the downward pass.

\begin{figure*}
\centering
    \subfloat[][A 3-level quadtree.]{\includegraphics[width=0.35\textwidth]{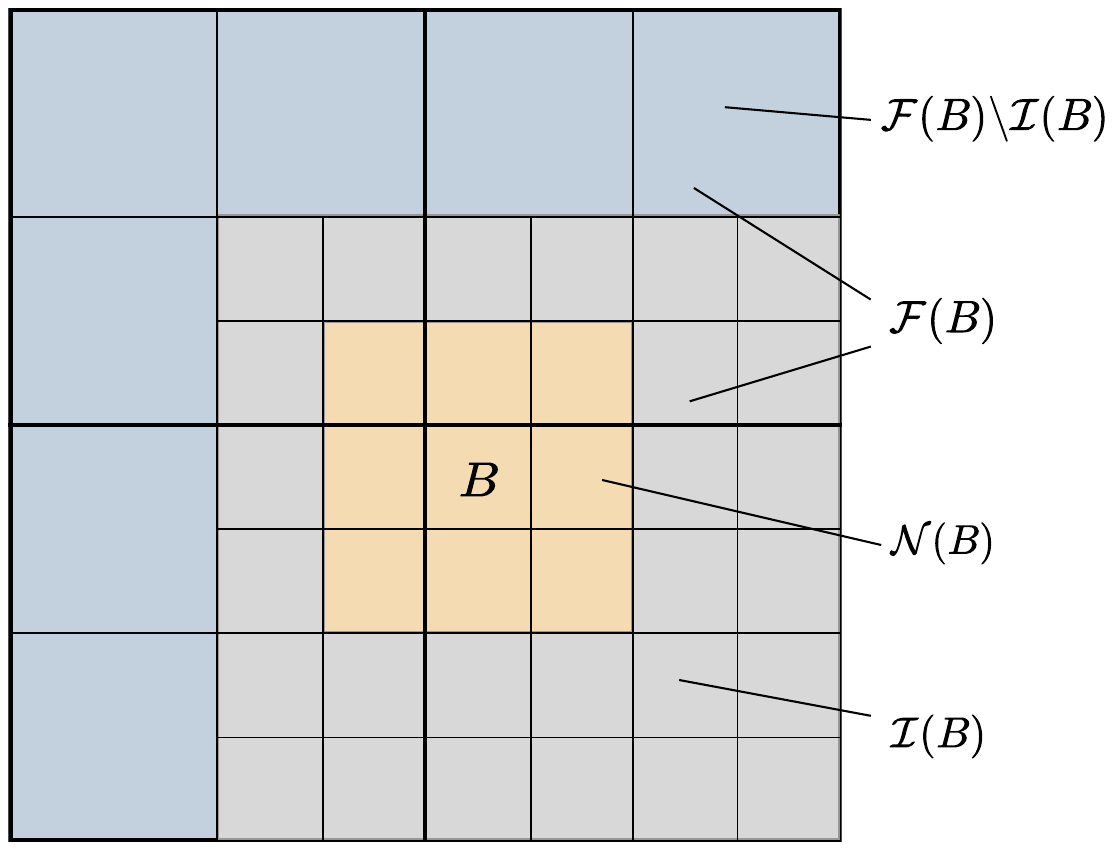}
        \label{fig:near_far_decomp}}
    \subfloat[][Sketch of FMM algorithm using a binary tree.]{\includegraphics[width=0.6\textwidth]{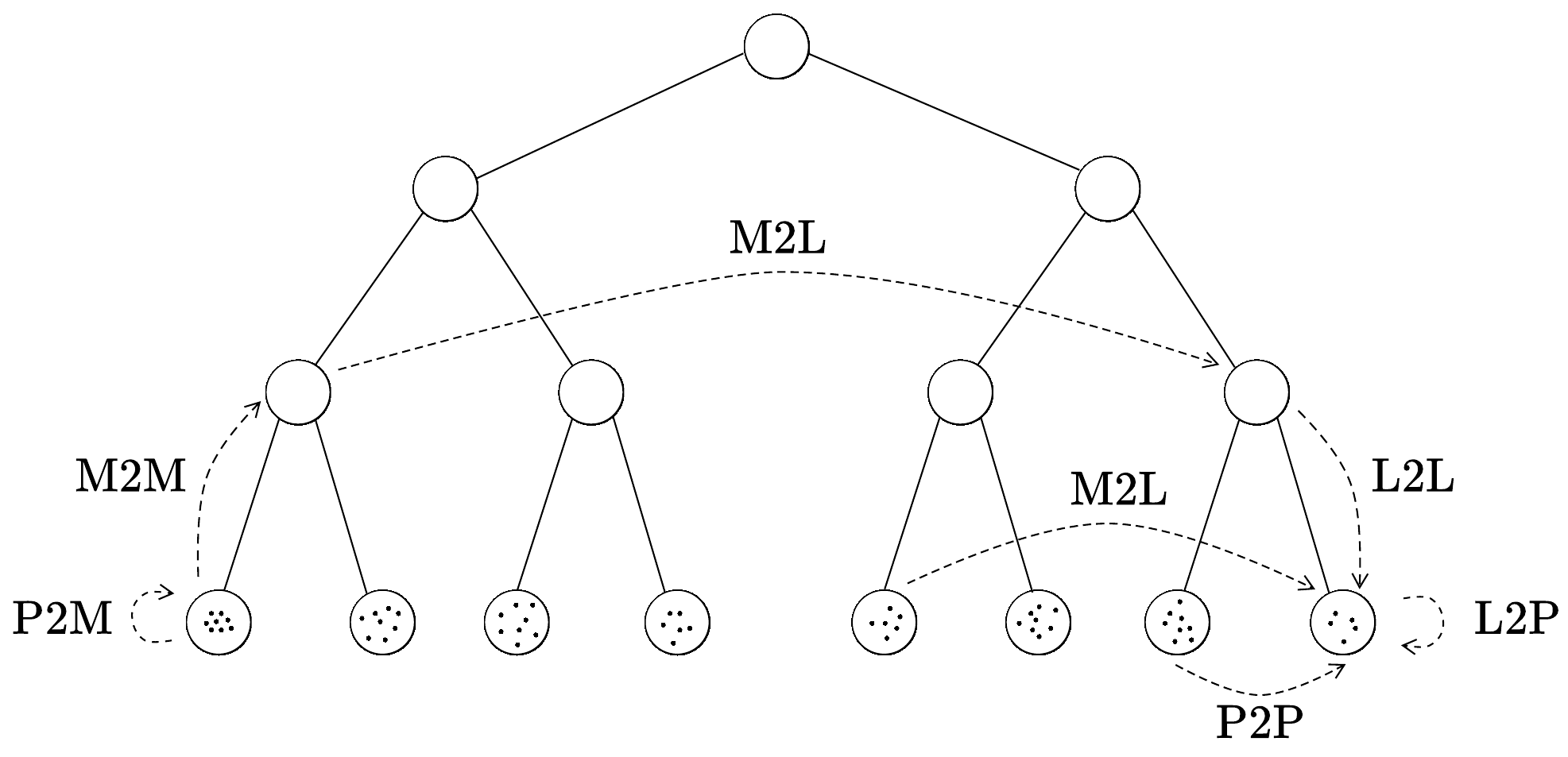}
        \label{fig:fmm_sketch}}\\
    \subfloat[][]{\includegraphics[width=0.4\textwidth]{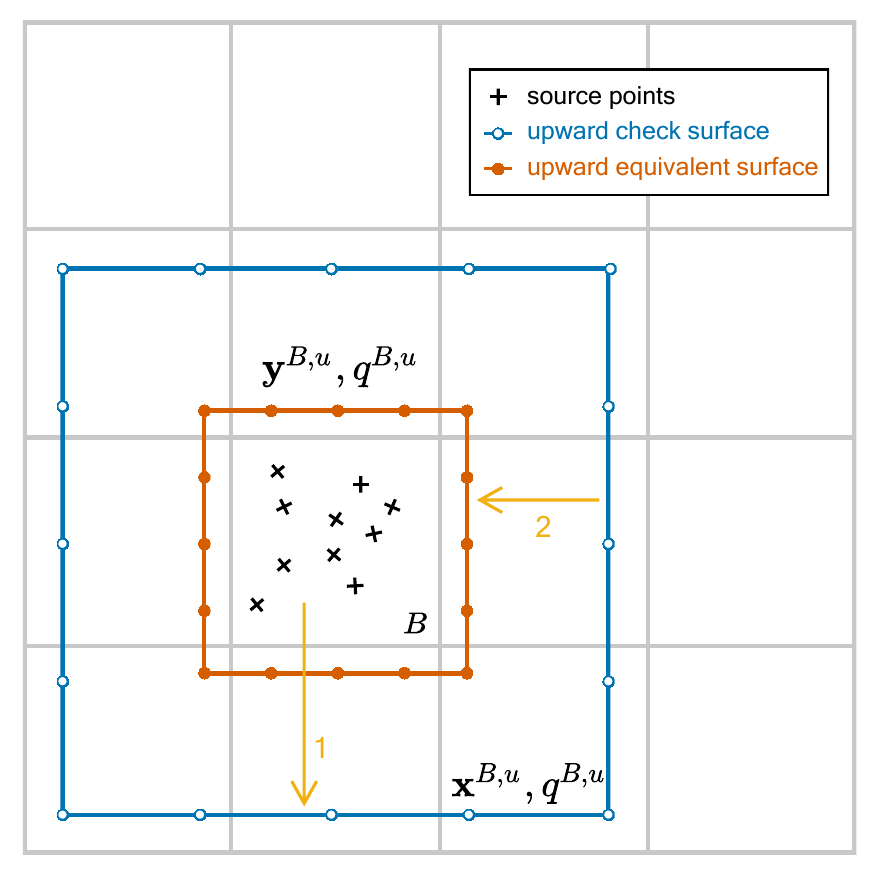}
        \label{fig:multipole}}
   \subfloat[][]{\includegraphics[width=0.4\textwidth]{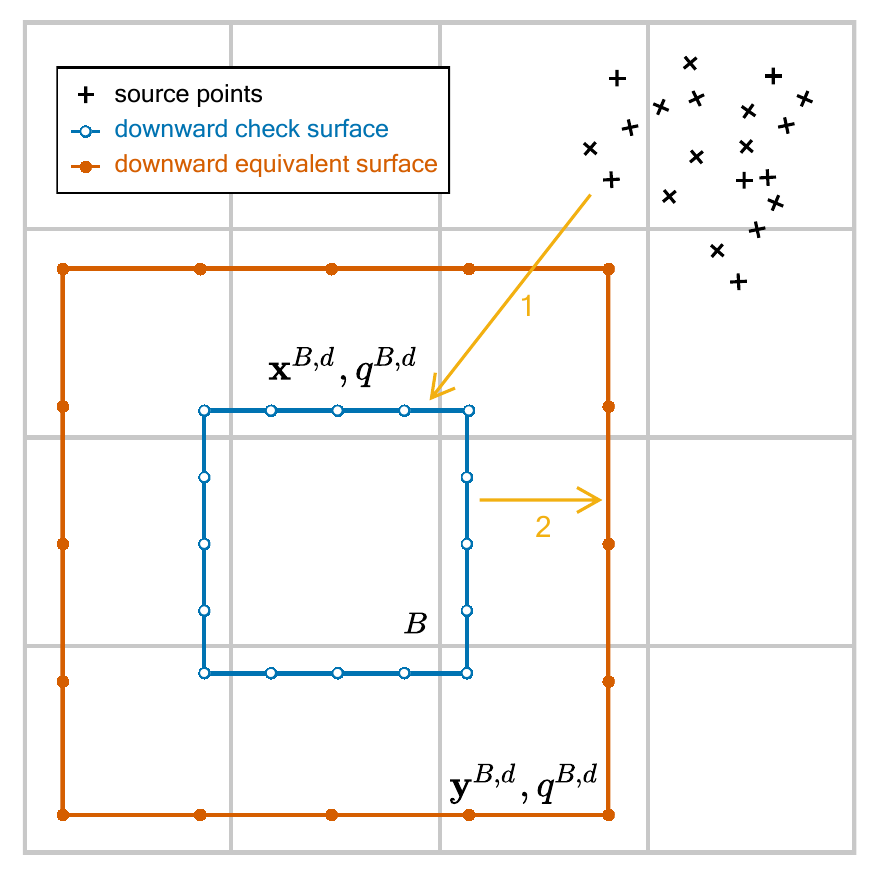}
        \label{fig:local}}\\
\subfloat[][]{\includegraphics[width=\textwidth]{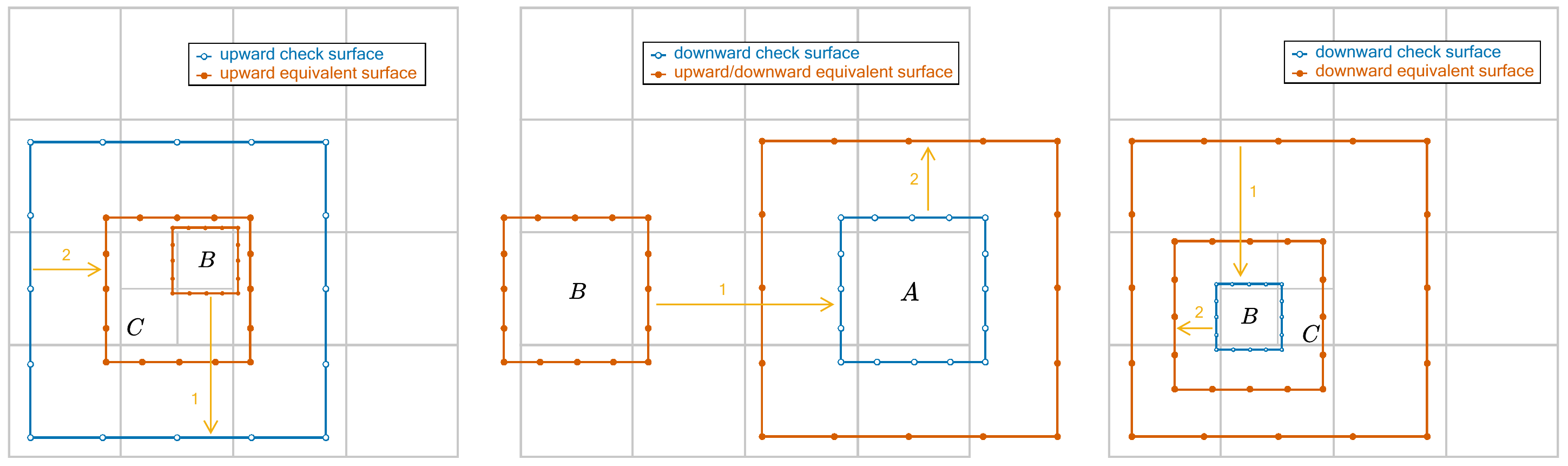}
        \label{fig:translations}}
    \caption{Illustrations of the \fmm algorithm.
    \textbf{c},\textbf{d}, Multipole and local expansion in \kifmm.
    \textbf{e}, M2M (left), M2L (middle) and L2L (right) operators in \kifmm. Node $C$ is the parent of $B$, and node $A$ is in the interaction list of $B$.
    }
\end{figure*}

The original Exafmm \cite{yokota2012tuned,yokota2013fmm} implements the classical \fmm based on dual tree traversal and focuses on low-accuracy optimizations.
Recently, Exafmm received a major update to adopt \kifmm due to its great extensibility.
Its current generation, Exafmm-t \cite{Wang2021}, offers highly optimized \kifmm operators, allows pre-computing and caching invariant matrices and more importantly, provides a high-level Python interface to reach a broader audience.
}

\section{Data availability}
We deposited the meshes and \texttt{pqr} files on the Zenodo service: \href{http://doi.org/10.5281/zenodo.4568768}{doi:10.5281/zenodo.4568768}.
The raw and secondary data for all results are available in the archival deposit of our paper’s GitHub repository: \href{http://doi.org/10.5281/zenodo.4568951}{doi:10.5281/zenodo.4568951}.

\section{Code availability}
Exafmm is available at \href{https://github.com/exafmm/exafmm-t}{https://github.com/exafmm/exafmm-t} under the BSD 3 license.
Bempp-cl is available at \href{https://github.com/bempp/bempp-cl}{https://github.com/bempp/bempp-cl} under the MIT license.
The scripts for plotting and rerunning our experiments are available in the archival deposit of our paper’s GitHub repository: \href{http://doi.org/10.5281/zenodo.4568951}{doi:10.5281/zenodo.4568951}.

\bibliography{./reference}{}
\bibliographystyle{elsarticle-num}

\section*{Acknowledgments}
We thank Dr. Sergio Pantano for providing us with the parameterized structure of the Zika virus capsid.
CDC acknowledges support by ANID (Agencia Nacional de Investigaci\'{o}n y Desarrollo) through PIA/APOYO AFB180002.
TB was supported by Engineering and Physical Sciences Research Council Grant EP/V001531/1.
LAB acknowledges funding from the National Science Foundation via award \#1747669.

\section*{Author contributions}
LAB and TB conceived this project. 
TW wrote the version of the Exafmm code used in this work, the Python bindings and the Bempp integration. 
TB gave technical support on Bempp usage and wrote code to aid the integration.
CDC gave conceptual advice and helped set up computational experiments.
TW ran the calculations and prepared the figures.
TW, CDC, TB, and LAB discussed and guided the conduct of the research, and interpreted the results.
TW wrote the first draft of the manuscript, and all authors contributed materially to the writing and revising.
LAB guided the data management and guarantees the preservation of the full research compendium for this work. 
All authors confirm that the figures and conclusions accurately reflect the research.

\section*{Competing interests}
The authors declare no competing interests.

\end{document}


\maketitle

\section*{Agreement of solvation energy calculations with APBS}
We build more confidence on Bempp-Exafmm by comparing our results with the finite-difference version of APBS (version 1.5), using 5PTI and 8 other molecules.
We prepared the molecular structures using the \texttt{charmm} force field.
For each molecule, we created three successive meshes to study the convergence with each software.
For Bempp-Exafmm, we used \texttt{Nanoshaper} to generate three surface meshes with the element density set to 4, 8, 16 ${\si{\angstrom}}^{-2}$, respectively.
In APBS, we controlled the multigrid parameters to ensure that three grid spacings are refined with a constant factor of 2.
Table \ref{tab:APBS_mesh} summarizes the mesh sizes of the molecules used in the convergence study of APBS and Bempp.
The former uses Cartesian grids while the latter uses surface meshes generated by \texttt{Nanoshaper}.
Other parameters are the same as our previous grid-convergence study, as listed in Table 1 in our paper.

Table \ref{tab:APBS_result} presents the convergence results of both codes using a set of molecules.
Bempp used the derivative formulation throughout.
In each case, we obtained an extrapolated value of the solvation energy of each molecule, which serves as the reference value to estimate the discretization error with each software.
As expected, the solvation energy computed from Bempp converges at the rate of $\mathcal{O}(N^{-1})$, where $N$ is the number of elements.
With APBS, the average observed order of convergence over all cases is 1.25, with respect to the grid spacing $h$; other studies \cite{CooperBardhanBarba2014,GengKrasny2013} have reported a similar rate. 
Our $\mathcal{O}(N^{-1})$ convergence corresponds to $\mathcal{O}(h^{-2})$ convergence in grid size, but for boundary elements (surface triangles) it is more natural to look at convergence with respect to $N$.
The difference in the extrapolated solvation energies obtained with the two codes varies between 0.8\% and 1.8\% across all cases.
A discrepancy is expected because (1) Bempp-Exafmm and APBS use different geometrical representations for the molecular surface, (2) the atomic charges in the molecule are mapped to grid points through interpolation in APBS, while Bempp treats them analytically and (3) the far-field boundary condition is exactly imposed in Bempp.
Furthermore, as Oberkampf and Roy \cite{oberkampf_roy_2010} remark on code verification, this discrepancy could be due to model differences since Bempp and APBS use different discrete equations.
We should only expect the same results when the same exact algorithm is employed.
Due to the same reasons, Geng and Krasny \cite{GengKrasny2013} observed a 1\% discrepancy in the solvation energy when comparing between TABI and APBS using 1A63 (Table 4).
Even among grid-based codes, discrepancies of such magnitude have been reported.
For example, Table IX in Geng et al. \cite{gengTreatmentChargeSingularities2007} reported a 2\% difference between MIBPB and APBS in results for protein 1SH1, using a grid spacing of 0.25A.

\bigskip

Readers can find the input files and results in this folder in the manuscript repository: \url{https://github.com/barbagroup/bempp_exafmm_paper/tree/master/repro-pack/runs/APBS_result}.

Readers can also find a Jupyter Notebook with the post-processing of results here: \url{https://github.com/barbagroup/bempp_exafmm_paper/blob/master/repro-pack/notebooks/compare_with_other_software.ipynb}.

\begin{table}[]
  \centering
  \begin{tabular}{c|cc|cc}
       & \multicolumn{2}{c|}{APBS}                     & \multicolumn{2}{c}{Bempp}                   \\
  ID   & $N_{grid}$ & $h_x, h_y, h_z (\si{\angstrom})$ & $N_{elem}$ & density $(1/\si{\angstrom}^2)$ \\ \hline
  1AJJ & $65^3$     & 0.65, 0.71, 0.79                 & 8,544      & 4                              \\
       & $129^3$    & 0.33, 0.36, 0.40                 & 17,200     & 8                              \\
       & $257^3$    & 0.17, 0.18, 0.20                 & 34,468     & 16                             \\ \hline
  1VJW & $97^3$     & 0.46, 0.56, 0.51                 & 11,544     & 4                              \\
       & $193^3$    & 0.23, 0.28, 0.25                 & 23,296     & 8                              \\
       & $385^3$    & 0.12, 0.14, 0.13                 & 46,660     & 16                             \\ \hline
  5PTI & $97^3$     & 0.46, 0.56, 0.51                 & 12,512     & 4                              \\
       & $193^3$    & 0.23, 0.28, 0.25                 & 25,204     & 8                              \\
       & $385^3$    & 0.12, 0.14, 0.13                 & 50,596     & 16                             \\ \hline
  1R69 & $97^3$     & 0.59, 0.54, 0.51                 & 12,004     & 4                              \\
       & $193^3$    & 0.30, 0.27, 0.25                 & 24,216     & 8                              \\
       & $385^3$    & 0.15, 0.14, 0.13                 & 48,648     & 16                             \\ \hline
  1A2S & $65^3$     & 0.92, 0.87, 1.01                 & 17,784     & 4                              \\
       & $129^3$    & 0.46, 0.44, 0.51                 & 35,732     & 8                              \\
       & $257^3$    & 0.23, 0.22, 0.25                 & 71,648     & 16                             \\ \hline
  1SVR & $97^3$     & 0.62, 0.68, 0.70                 & 18,644     & 4                              \\
       & $193^3$    & 0.31, 0.34, 0.35                 & 37,348     & 8                              \\
       & $385^3$    & 0.16, 0.17, 0.18                 & 75,004     & 16                             \\ \hline
  1A63 & $129^3$    & 0.82, 0.50, 0.51                 & 27,996     & 4                              \\
       & $257^3$    & 0.41, 0.25, 0.25                 & 56,356     & 8                              \\
       & $513^3$    & 0.21, 0.13, 0.13                 & 113,124    & 16                             \\ \hline
  1A7M & $129^3$    & 0.52, 0.77, 0.63                 & 30,576     & 4                              \\
       & $257^3$    & 0.26, 0.39, 0.32                 & 61,652     & 8                              \\
       & $513^3$    & 0.13, 0.19, 0.16                 & 123,776    & 16                             \\ \hline
  1F6W & $129^3$    & 0.98, 0.79, 0.91                 & 77,464     & 4                              \\
       & $257^3$    & 0.49, 0.40, 0.46                 & 156,140    & 8                              \\
       & $513^3$    & 0.25, 0.20, 0.23                 & 313,692    & 16                            
  \end{tabular}
  \caption{Mesh sizes of the molecules used in the convergence study of APBS and Bempp.
  $h_x$, $h_y$ and $h_z$ are the grid spacings in $x$, $y$, $z$ direction.}
  \label{tab:APBS_mesh}
\end{table}

\begin{landscape}
\begin{table}[]
    \centering
    \resizebox{1.4\textheight}{!}{%
    \begin{tabular}{cc|ccc|cc|ccc|cc}
      &   & \multicolumn{5}{c|}{APBS}                                                                    & \multicolumn{5}{c}{Bempp}                                                                   \\
      &   & \multicolumn{3}{c|}{Error}     & \multirow{2}{*}{$\Delta G_{solv}$} & \multirow{2}{*}{Order$(1/h)$} & \multicolumn{3}{c|}{Error}     & \multirow{2}{*}{$\Delta G_{solv}$} & \multirow{2}{*}{Order$(1/N)$} \\
    ID   & $N_{atoms}$ & coarse   & medium   & fine     &                                    &                        & coarse   & medium   & fine     &                                    &                        \\ \hline
    1AJJ & 513 & 4.40e-02 & 1.33e-02 & 4.00e-03 & -266.15                            & 1.73                   & 2.75e-02 & 1.12e-02 & 4.59e-03 & -268.67                            & 1.29                   \\
    1VJW & 826 & 3.09e-02 & 1.45e-02 & 6.77e-03 & -297.06                            & 1.09                   & 4.92e-02 & 2.23e-02 & 1.01e-02 & -302.50                            & 1.14                   \\
    5PTI & 892 & 3.68e-02 & 1.38e-02 & 5.14e-03 & -311.69                            & 1.42                   & 5.12e-02 & 2.23e-02 & 9.67e-03 & -314.34                            & 1.20                   \\
    1R69 & 997 & 4.31e-02 & 2.10e-02 & 1.02e-02 & -261.02                            & 1.04                   & 5.09e-02 & 2.34e-02 & 1.08e-02 & -265.02                            & 1.12                   \\
    1A2S & 1272 & 5.25e-02 & 2.40e-02 & 1.10e-02 & -456.56                            & 1.13                   & 4.21e-02 & 1.93e-02 & 8.86e-03 & -461.25                            & 1.12                   \\
    1SVR & 1433 & 5.28e-02 & 2.21e-02 & 9.28e-03 & -393.45                            & 1.25                   & 6.48e-02 & 2.89e-02 & 1.29e-02 & -398.83                            & 1.16                   \\
    1A63 & 2065 & 4.51e-02 & 1.88e-02 & 7.82e-03 & -559.39                            & 1.26                   & 5.82e-02 & 2.60e-02 & 1.16e-02 & -567.24                            & 1.16                   \\
    1A7M & 2804 & 4.13e-02 & 1.88e-02 & 8.53e-03 & -524.29                            & 1.14                   & 4.93e-02 & 2.24e-02 & 1.02e-02 & -531.48                            & 1.14                   \\
    1F6W & 8247 & 6.45e-02 & 2.78e-02 & 1.20e-02 & -1277.51                           & 1.21                   & 4.18e-02 & 1.80e-02 & 7.76e-03 & -1301.08                           & 1.22                     
    \end{tabular}
    }
    \caption{Convergence results of the solvation energy of 9 molecules using APBS and Bempp with derivative formulation.
    The error is with respect to an extrapolated value of the solvation energy using Richardson extrapolation.
    The solvation energy $\Delta G_{solv}$ is in units of kcal/mol.
    The observed order of convergence is with respect to the grid spacing $h$ for APBS (volumetric-based solver) and with respect to the number of elements $N$ for Bempp (boundary-element solver).}
    \label{tab:APBS_result}
\end{table}
\end{landscape}

\section*{Binding energy calculations}

Table \ref{tab:solv_bind} lists the solvation energies of 9 Barnase-Barstar complexes and their components.
The data in the first 4 columns are directly from the supporting information of Nguyen \emph{et al.} \cite{nguyenAccurateRobustReliable2017}.

\begin{table*}[]
     \centering
     \begin{tabular}{cc|c|c|c|c|c}
                           &         & MIBPB   & DELPHI  & PBSA    & APBS    & Bempp   \\ \hline
     \multirow{3}{*}{1b27} & complex & -2249.3 & -2275.1 & -2276.6 & -2278.3 & -2272.8 \\
                           & barnase & -1271.9 & -1280.8 & -1281.2 & -1287.5 & -1278.8 \\
                           & barstar & -1573.2 & -1582.4 & -1583.8 & -1596.9 & -1585.1 \\ \hline
     \multirow{3}{*}{1b2s} & complex & -2479.2 & -2484.1 & -2484.9 & -2486.6 & -2484.6 \\
                           & barnase & -1215.1 & -1222.8 & -1223.5 & -1228.0 & -1220.1 \\
                           & barstar & -1623.4 & -1628.1 & -1628.4 & -1642.2 & -1629.8 \\ \hline
     \multirow{3}{*}{1b2u} & complex & -2395.2 & -2402.4 & -2404.4 & -2401.7 & -2403.4 \\
                           & barnase & -1260.6 & -1266.6 & -1267.3 & -1272.8 & -1263.1 \\
                           & barstar & -1431.9 & -1442.6 & -1443.8 & -1454.7 & -1443.5 \\ \hline
     \multirow{3}{*}{1b3s} & complex & -2284.8 & -2301.1 & -2301.7 & -2299.5 & -2298.2 \\
                           & barnase & -1294.6 & -1302.2 & -1302.4 & -1310.0 & -1299.1 \\
                           & barstar & -1588.0 & -1596.4 & -1597.2 & -1609.3 & -1599.0 \\ \hline
     \multirow{3}{*}{1x1u} & complex & -2259.8 & -2266.2 & -2267.2 & -2264.1 & -2266.1 \\
                           & barnase & -1305.8 & -1313.4 & -1313.7 & -1320.4 & -1310.8 \\
                           & barstar & -1580.4 & -1583.6 & -1585.1 & -1594.3 & -1585.8 \\ \hline
     \multirow{3}{*}{1x1w} & complex & -2074.3 & -2085.6 & -2086.0 & -2085.8 & -2082.5 \\
                           & barnase & -1250.9 & -1254.8 & -1254.7 & -1261.7 & -1250.5 \\
                           & barstar & -1372.4 & -1378.8 & -1379.0 & -1388.0 & -1380.0 \\ \hline
     \multirow{3}{*}{1x1x} & complex & -2183.0 & -2194.5 & -2196.1 & -2196.6 & -2191.9 \\
                           & barnase & -1284.0 & -1296.2 & -1296.7 & -1305.0 & -1293.1 \\
                           & barstar & -1401.0 & -1407.8 & -1408.6 & -1416.6 & -1407.8 \\ \hline
     \multirow{3}{*}{1x1y} & complex & -2063.9 & -2072.5 & -2073.1 & -2076.3 & -2077.0 \\
                           & barnase & -1260.6 & -1266.6 & -1267.0 & -1273.6 & -1263.9 \\
                           & barstar & -1309.8 & -1321.1 & -1321.7 & -1331.2 & -1321.6 \\ \hline
     \multirow{3}{*}{2za4} & complex & -2186.9 & -2199.8 & -2201.2 & -2190.7 & -2200.2 \\
                           & barnase & -1190.2 & -1200.3 & -1201.5 & -1207.2 & -1197.9 \\
                           & barstar & -1369.4 & -1380.4 & -1381.5 & -1393.5 & -1381.2
     \end{tabular}
     \caption{Solvation energies of 9 Barnase-Barstar complexes and their components, using various PB solvers.
     Energies are in units of kcal/mol.
     \label{tab:solv_bind}
     }
\end{table*}

 Table \ref{tab:coul_bind} summarizes $\Delta G_{coul}$, the difference in Coulombic energy, for the 9 complexes.

\begin{table}[]
     \centering
     \begin{tabular}{c|cc}
          & $\Delta G_{coul}$ [kcal/mol] &  \\ \hline
     1b27 & -499.0       &  \\
     1b2s & -289.7       &  \\
     1b2u & -221.9       &  \\
     1b3s & -546.8       &  \\
     1x1u & -550.4       &  \\
     1x1w & -448.7       &  \\
     1x1x & -390.4       &  \\
     1x1y & -415.6       & 
     \end{tabular}
     \caption{$\Delta G_{coul}$ for 9 different Barnase-Barstar complexes.}
     \label{tab:coul_bind}
\end{table}     

\bibliography{./reference}{}
\bibliographystyle{elsarticle-num}